\documentclass[
reprint,
superscriptaddress,
amsmath,amssymb,
aps,
pra,
floatfix,
]{revtex4-2}

\usepackage{graphicx}% Include figure files
\usepackage{dcolumn}% Align table columns on decimal point
\usepackage{bm}% bold math
\usepackage{braket}
\usepackage{float}
\usepackage{xcolor}
\usepackage{hyperref}% add hypertext capabilities
\usepackage{cleveref}
\usepackage{physics}
\usepackage{cancel}
\usepackage{xcolor}
\usepackage{dsfont}
\usepackage{listings} %for writing code snippets
\definecolor{codegreen}{rgb}{0,0.6,0}
\definecolor{codegray}{rgb}{0.5,0.5,0.5}
\definecolor{codepurple}{rgb}{0.58,0,0.82}
\definecolor{backcolour}{rgb}{0.95,0.95,0.92}

\lstdefinestyle{mystyle}{
    backgroundcolor=\color{backcolour},   
    commentstyle=\color{codegreen},
    keywordstyle=\color{magenta},
    numberstyle=\tiny\color{codegray},
    stringstyle=\color{codepurple},
    basicstyle=\ttfamily\footnotesize,
    breakatwhitespace=false,         
    breaklines=true,                 
    captionpos=b,                    
    keepspaces=true,                 
    numbers=left,                    
    numbersep=5pt,                  
    showspaces=false,                
    showstringspaces=false,
    showtabs=false,                  
    tabsize=2
}

\usepackage[mathlines]{lineno}% Enable numbering of text and display math
\def\[#1\]{\begin{align}#1\end{align}}

\begin{document}

%\title{The implementation of fast simulations of continuous-variable circuits using linear combinations of coherent states}
%\title{Numerical simulations of continuous-variable circuits using linear combinations of coherent states}
\title{Fast simulations of continuous-variable circuits using the coherent state decomposition}
\author{Olga Solodovnikova}
\email{olgasol@dtu.dk}
\author{Ulrik L. Andersen}
\email{ulrik.andersen@fysik.dtu.dk}
\author{Jonas S. Neergaard-Nielsen}
\email{jsne@fysik.dtu.dk}
\affiliation{Center for Macroscopic Quantum States (bigQ), Department of Physics, Technical University of Denmark, Building 307, Fysikvej, 2800 Kgs. Lyngby, Denmark}
\date{August 8, 2025}

\begin{abstract}
We present \texttt{lcg\_plus}, an open-source Python library for the simulation of continuous-variable quantum circuits with both generaldyne and photon-number-resolving detector capabilities. Our framework merges the linear combination of Gaussians methodology with the coherent state decomposition of arbitrary non-Gaussian states, forming a bridge between the Gaussian and Fock basis representations. By tracking the Wigner function, we can simulate the action of Gaussian channels and measurements on multi-mode systems in a fast and accurate numerical framework. The calculation of the quality measures of quantum states is convenient in this formalism, and we derive expressions for the analytical gradients of these measures with respect to parameterized circuit elements. We demonstrate the utility of this methodology by optimizing the heralded preparation of a qunaught state, a crucial component for building a fault-tolerant photonic quantum computer, with a Gaussian Boson sampling circuit containing inefficient components.

\end{abstract}

\maketitle
\section{Introduction}

Photonic systems are excellent platforms for encoding and processing quantum information \cite{braunstein_quantum_2005, menicucci_fault-tolerant_2014, andersen_hybrid_2015}, thanks to their low-noise properties  and networking capability. These features make them attractive for a wide range of quantum technologies such as quantum computing \cite{psiquantum_team_manufacturable_2025, aghaee_rad_scaling_2025}, communication \cite{lo_secure_2014,azuma_all-photonic_2015, pirandola_advances_2020, usenko_continuous-variable_2025}, and metrology \cite{mitchell_super-resolving_2004,giovannetti_quantum_2006,nielsen_deterministic_2023}. Quantum information can be encoded in discrete variables (DV), such as the polarization of a single photon, or in continuous variables (CV) - the quadratures of the electromagnetic field. 

The preparation of high-quality non-Gaussian states is crucial for many applications within quantum information processing, such as entanglement distillation \cite{eisert_distilling_2002} and loophole-free tests of Bell inequalities \cite{garcia-patron_proposal_2004}. In particular for quantum computing, single photons must be prepared and entangled in the fusion-based quantum computation model \cite{bartolucci_fusion-based_2023}. In CV, the qubit is encoded in the infinite dimensional Hilbert space of a bosonic system that exhibits translational symmetry \cite{gottesman_encoding_2001}, or rotational symmetry \cite{grimsmo_quantum_2020} in phase space. A reliable source of Gottesman-Kitaev-Preskill (GKP) states will enable fault-tolerant universal quantum computing \cite{menicucci_universal_2006, baragiola_all-gaussian_2019}. However, it is still an outstanding challenge to produce GKP states at high enough rate and quality required for fault-tolerant quantum computing with current technology. 

In optics, the lack of a strong nonlinearity, such as the Kerr non-linearity, necessitates the use of photon detectors to "de-Gaussify" a multi-mode entangled Gaussian state \cite{su_conversion_2019,eaton_non-gaussian_2019, takase_generation_2021,takase_generation_2024} and non-Gaussian states have been generated experimentally in traveling light \cite{ourjoumtsev_generating_2006, ourjoumtsev_generation_2007, konno_logical_2024, endo_high-rate_2025, larsen_integrated_2025}. However, the omnipresent photon loss and detector inefficiencies are detrimental to the Wigner-negativities in the quasi-probability distribution in phase space, a key feature of non-Gaussian states. 

% Alternatively, squeezed Schrödinger cat states can be bred together using Gaussian operations, homodyne detection and a feedforward displacement operation to produce a GKP state \cite{vasconcelos_all-optical_2010, weigand_generating_2018}. 

% The generation of the input cat states, however, still relies on using photon detectors to detect relatively large photon numbers .

% For the  encoding, a measurement-based quantum computing model can be employed to perform the required Clifford gates enabling fault-tolerant universal quantum computing.

To design and control quantum state preparation protocols under realistic conditions, accurate and efficient simulation tools are required. This is especially important in setups involving Gaussian operations combined with photon detections which is a class of problems known as Gaussian Boson sampling (GBS) \cite{hamilton_gaussian_2017}. Quantum advantage with GBS has been claimed experimentally \cite{zhong_quantum_2020, madsen_quantum_2022}, however the photon loss levels in the experiments proved to be too high for actual quantum advantage \cite{oh_classical_2024}. Photon loss is a major obstacle for practical applications of photonics, and must therefore be accounted for when designing protocols.

There are mainly two ways to simulate CV circuits in optics. All Gaussian operations can be efficiently simulated using the first two moments of the position and momentum operators. Alternatively, operations can be performed in the photon numbers basis, which grows exponentially large with the number of modes. Several simulation frameworks and numerical libraries have been developed over the last years specifically for the CV formalism,  such as 
% \texttt{qutip} \cite{johansson_qutip_2012}, 
\texttt{strawberryfields} \cite{killoran_strawberry_2019}, \texttt{thewalrus} \cite{gupt_walrus_2019}, \texttt{MrMustard} \cite{yao_riemannian_2024} and \texttt{piquasso} \cite{kolarovszki_piquasso_2025}. In most of these frameworks, Gaussian circuit elements such as squeezers, beam splitters, phase shifters, displacements, and homodyne/heterodyne detectors are implemented via transformations of the covariance matrix and displacement vector. As soon as a non-Gaussian element appears, such as a Kerr gate, cubic phase gate or the photon detector, the first two moments of the quadrature operators no longer fully characterize the state. Upon encountering a non-Gaussian state or element, the state representation is switched to the Fock basis. Since the basis is infinite, a suitable photon number cutoff must be chosen, which can lead to numerical errors \cite{provaznik_taming_2022} if done improperly. Once the switch to the Fock basis is made, the convenient transformation rules for Gaussian states are no longer applicable, and the simulation of high energy multi-mode states becomes cumbersome, especially for the simulation of the density operator. Tensor-network based methods for CV systems have also started to appear, in which absence of entanglement is leveraged by constructing matrix product states, which use smaller dimensions in Fock space \cite{oh_classical_2024,vinther_variational_2024} and in phase space
\cite{michelsen_functional_2025}, however this is only beneficial when the entanglement is low. 

An alternative approach is to represent non-Gaussian states and operations with linear combinations of Gaussians (LCoG) \cite{bourassa_fast_2021}. Here, the convenience of the Gaussian formalism is retained by keeping track of a weighted sum of Gaussians, represented by several covariance matrices, displacement vectors, and weights.  This framework is implemented in the \texttt{bosonic} backend of \texttt{strawberryfields}, and is mainly used for fast simulation of bosonic qubits undergoing multi-mode Gaussian channels. Recently, efficient frameworks for simulating superpositions of Gaussian states undergoing linear optical circuits have also been proposed \cite{marshall_simulation_2023, dias_classical_2024}. While an approximation of Fock states with LCoG has been proposed in \cite{bourassa_fast_2021}, they are numerically unstable, and therefore photon-number-resolving detectors (PNRDs) are not available in the \texttt{bosonic} backend. 

In this contribution, we expand upon the LCoG framework by combining it with a new, numerically stable way to approximate arbitrary superpositions of Fock states by using the coherent state decomposition \cite{marshall_simulation_2023}. 
With this addition, the LCoG framework becomes very general, and we are able to simulate Gaussian channels on any non-Gaussian quantum state, and use both generaldyne and photon-number-resolving detectors. The complexity of the simulation framework can be related to the stellar rank \cite{chabaud_stellar_2020,chabaud_classical_2021} of the system, capturing the minimal number of non-Gaussian resources.
As a demonstration, we simulate and optimize non-Gaussian state preparation circuits consisting of Gaussian operations and PNRDs, as well as include Gaussian channels that model photon loss and detector inefficiencies in a fast and accurate numerical framework. Figures of merit, such as the fidelity, the effective squeezing \cite{duivenvoorden_single-mode_2017}, and the nonlinear GKP squeezing \cite{marek_ground_2024} can be calculated straightforwardly from LCoG. Additionally, the analytical gradients of these quantities with respect to the parameters of a GBS circuit can be derived and used to enhance the optimization. 
%This has, to our knowledge, not previously been done for linear combinations of Gaussians. 

We implement these features, in addition to other functionalities such as pseudo-PNRDs and the ability to reduce the number of Gaussians required to represent a single-mode non-Gaussian state, enabling a speedup for certain circuit topologies. 

The paper is structured as follows; in Sec.\ \ref{sec:background}, we review the Gaussian CV formalism, and introduce the LCoG formalism and the coherent state decomposition. In Sec.\ \ref{sec:lcog_simulation}, we describe the numerical framework for simulating and optimizing GBS devices for non-Gaussian (GKP) state preparation, derive the gradient calculation, and discuss strategies for reducing the number of Gaussians. In Sec. \ref{sec:results}, the results of the optimization of GBS circuits for qunaught state preparation are presented. Lastly, in Sec.\ \ref{sec:summary} we suggest further applications for our framework. 
\section{Background}
\label{sec:background}
We consider $N$ modes of the electromagnetic field represented by a vector of the quadrature operators  $\hat{\vb*{q}}=(\hat{x}_{1},\hat{p}_{1},\dots,\hat{x}_{N},\hat{p}_{N})^T$, where $\hat{x}_i$ is the position and $\hat{p}_i$ is the momentum quadrature of mode $i$. They satisfy the following commutator relations $[\hat{q}_{i},\hat{q}_{j}]=i\hbar\Omega_{ij}$ where $\Omega_{ij}$ is an element of the $2N\times  2N$ symplectic form, $\vb{\Omega}=\bigoplus_{i=1}^N\big(\begin{smallmatrix}
0 & 1 \\-1 & 0
\end{smallmatrix}\big)$. 
\subsection{Gaussian Formalism}
\label{sec:gauss_formalism}
In the following, we summarize the Wigner characteristic function formalism for Gaussian transformations, Gaussian states, Gaussian channels, and Gaussian measurements. 
\subsubsection*{Quasi-probability distributions}
A multimode bosonic system $\hat{\rho}$ is fully characterized by its quasi-probability distribution in phase-space, the Wigner characteristic function $\chi(\vb*{\alpha})$;
\[
 \chi_{\hat{\rho}}(\vb*{\alpha})=\langle\hat{D}(\vb*{\alpha})\rangle = &\Tr[\hat{\rho}\hat{D}(\vb*{\alpha})],
\]
where $\hat{D}(\vb*{\alpha})=\exp[i\sqrt{\frac{2}{\hbar}}\hat{\vb*{q}}^T\vb*{\Omega}\vb*{\alpha}]$ is the (symmetrically ordered) Weyl operator, which induces displacements in phase space, $\hat{D}^{\dagger}(\vb*{\alpha})\hat{\vb*{q}}\hat{D}(\vb*{\alpha})=\hat{\vb*{q}}+\sqrt{ 2\hbar }\vb*{\alpha}$. The multivariate coordinate $\vb*{\alpha}$ is a vector over the real and imaginary parts of the displacement, e.g. for a single mode $\vb*{\alpha}=(\Re\alpha, \Im\alpha)^T$. The Wigner function is related to the characteristic function via a Fourier transform,
\[
W_{\hat{\rho}}(\vb*{q})=\int_{\mathds{R}^{2N}} \frac{\dd^{2N}{\vb*{\alpha}}}{(2\pi^2 \hbar)^{N}}e^{-i\sqrt{\frac{2}{\hbar}}\vb*{q}^T\vb*{\Omega}\vb*{\alpha}}\chi_{\hat{\rho}}(\vb*{\alpha}),
\]
and is an equivalent representation of the state $\hat{\rho}$. If $W_{\hat{\rho}}(\vb*{q})$ is a LCoG, then $\chi_{\hat{\rho}}(\vb*{\Lambda})$ is a LCoG.
%of the Fourier transformed-Gaussians. 
\subsubsection*{Symplectic transformations}
\label{sec:sympl}
Most of the commonly available components in quantum optics such as squeezers, phase shifters, and beam splitters are Gaussian.
A Gaussian (Bogoliubov) transformation has a unitary operator $\hat{U}=\exp[-i\hat{H}/\hbar]$ where the Hamiltonian is at most quadratic in the bosonic or quadrature operators. In the Heisenberg picture, the quadrature operators transform via a symplectic map, $\hat{U}^\dagger \hat{\vb*{q}}\hat
U=\vb*{S}\hat{\vb*{q}}+\vb*{d}$, where $\vb*{S}$ is a symplectic matrix preserving the symplectic form, $\vb*{S}\vb*{\Omega}\vb*{S}^T=\vb*{\Omega}$, and $\vb*{d}$ is a vector of phase-space displacements. The symplectic matrices for various operators used in quantum optics can be found in several references \cite{weedbrook_gaussian_2012, brask_gaussian_2022}. Using the inverse correspondence rule, the coordinate of the Wigner function can simply be transformed to evolve the system, $W_{\hat{U}\hat{\rho}\hat{U}^{\dagger}}(\vb*{q})=W_{\hat{\rho}}(\vb*{S}^{-1}(\vb*{q}-\vb*{d}))$. 

\subsubsection*{Gaussian states}
\label{sec:Gaussian_states}
A general Gaussian state can be constructed by applying Gaussian transformations on thermal states, $\hat{U}\hat{\rho}_\text{th}\hat{U}^\dagger$. Gaussian states are fully described by their first and second quadrature moments; the displacement vector $\vb*{\mu}$ with elements $\mu_i=\langle \hat{q}_i \rangle$, and the covariance matrix $\vb*{\sigma}$ with elements $\sigma_{ij}=\frac{1}{2}\langle \{\hat{q}_i-\langle\hat{q}_i\rangle,\hat{q}_j-\langle\hat{q}_j\rangle\}\rangle$, where the curly brackets indicate the anti-commutator. The covariance matrix must be real, symmetric and positive definite, $\vb*{\sigma}>0$, and must satisfy the uncertainty relation $\vb*{\sigma}+i\frac{\hbar}{2}\vb*{\Omega}\geq 0$ for all valid quantum states \cite{simon_quantum-noise_1994}. The quasi-probability distribution of a Gaussian state is a multivariate Gaussian, 
\[
W_{\hat{U}\hat{\rho}_{\text{th}}\hat{U}^\dagger}(\vb*{q})=G_{\vb*{\mu},\vb*{\sigma}}(\vb*{q}),
\]
where the multivariate Gaussian function is defined as
\[
G_{\vb*{\mu},\vb*{\sigma}}(\vb*{q})=\frac{1}{\sqrt{ \det(2\pi \vb*{\sigma}) }}\exp\left[ -\frac{1}{2}(\vb*{q}-\vb*{\mu})^T \vb*{\sigma}^{-1}(\vb*{q}-\vb*{\mu}) \right].
\]
The covariance matrix can be decomposed via the Williamson decomposition \cite{williamson_algebraic_1936} to be
$\vb*{\sigma}=\vb*{S}\vb*{D}\vb*{S}^T$ where $\vb*{D}=\bigoplus_{i=1}^N\big(\begin{smallmatrix}
    \nu_i & 0 \\ 0 & \nu_i
\end{smallmatrix}\big)$ is a diagonal matrix containing the variances $\nu_i=\frac{\hbar}{2} ( 2\Bar{n}_i +1)$ of the thermal states with occupation numbers $\bar{n}_i$,  and $\vb*{S}$ is the symplectic matrix of the Gaussian unitary $\hat{U}$. $\vb*{\mu}=\vb*{d}$ is a vector of phase space displacements. 

Further Gaussian operations on the Gaussian state can be performed by transforming the initial displacement vector and the covariance matrix, $\vb*{\mu}\mapsto \vb*{S}\vb*{\mu}+\vb*{d}$ and $\vb*{\sigma}\mapsto \vb*{S}\vb*{\sigma}\vb*{S}^T$, or by using the inverse correspondence rule and transforming the coordinate of the Wigner function, as discussed before. 

The characteristic function \footnote{The displacement in the exponent is sometimes written as $-i(\vb*{\Omega}\vb*{\mu})^T\vb*{\alpha}$} of a Gaussian state is \cite{serafini_quantum_2017}
\[
\chi_{\hat{\rho}_G}(\vb*{\alpha}) &= \exp\left[-\frac{1}{2}\vb*{\alpha}^T\vb*{\Omega}\vb*{\sigma}\vb*{\Omega}^T\vb*{\alpha}+i\vb*{\mu}^T\vb*{\Omega}\vb*{\alpha}\right].
% &=G_{\vb*{0},\tilde{\vb*{\sigma}}}(\vb*{\Lambda})\exp[i\vb*{\phi}^T\vb*{\Lambda}]
\]% where $\vb*{\Omega}$ is the symplectic form, $\tilde{\vb*{\sigma}} = (\vb*{\Omega}\vb*{\sigma}\vb*{\Omega}^T)^{-1}$ and $\vb*{\phi}=\vb*{\Omega}\vb*{\mu}$.
\subsubsection*{Gaussian channels}
A channel is a linear, completely positive, trace-preserving (CPTP) map. A channel is Gaussian if it deterministically maps Gaussian states to Gaussian states. This encompasses many of the noise mechanisms in optical systems such as loss, gain and random Gaussian displacements. A Gaussian unitary transform is also a specific case of a Gaussian channel. These can be implemented on a Gaussian state by a simple transform of the first and second moments: $\vb*{\mu}\mapsto\vb*{X}\vb*{\mu}$ and $\vb*{\sigma}\mapsto \vb*{X}\vb*{\sigma}\vb*{X}^T+\vb*{Y}$. The $(\vb*{X},\vb*{Y})$ matrices must satisfy $\vb*{Y}+\frac{i\hbar}{2}\vb*{\Omega}\geq i \frac{\hbar}{2}\vb*{X}\vb*{\Omega}\vb*{X}^T$ \cite{serafini_quantum_2017}, and are listed in Table \ref{tab:G_channels} for some common Gaussian channels. 
\begin{table}[b]
    \centering
    \caption{$(\vb*{X}$, $\vb*{Y})$ matrices for the single-mode Gaussian channels: the loss/attenuation channel, the gain/amplification channel, the random Gaussian displacement channel, and the reversible Gaussian unitary channel. $0\leq\eta\leq1$ is the transmissivity of the loss channel and $\bar{n}\geq0$ is the average photon number the thermal noise ($\bar{n}=0$ for pure loss channel). $G\geq1$ is the gain.}
    \begin{ruledtabular}
    \begin{tabular}{ccc}
    %\hline
       Channel  & $\vb*{X}$ & $\vb*{Y}$  \\
       \colrule
       Loss/Attenuation & $\sqrt{\eta}\mathds{1}_2$ & $\frac{\hbar}{2}(1-\eta)(2\bar{n}+1) \mathds{1}_2$\\
       Gain/Amplification & $\sqrt{G}\mathds{1}_2$  & $\frac{\hbar}{2}(G -1) \mathds{1}_2$\\
       Random Gaussian Displacement & $\mathds{1}_{2}$ & $\vb*{W}$\\
       Gaussian Unitary & $\vb*{S}$ & $\vb*{0}$
       %\hline
    \end{tabular}
    \end{ruledtabular}
    
    \label{tab:G_channels}
\end{table}
\subsubsection*{Gaussian measurements} 
\label{sec:Gaussian_measurements}
A measurement of a quantum state $\hat{\rho}$ with a result $\vb*{m}$ occurs with probability,
\begin{equation}
    p(\vb*{m})=\Tr[\hat{\rho}\hat{\Pi}(\vb*{m})],
\end{equation}
where $\hat{\Pi}(\vb*{m})$ is the positive operator-valued measure (POVM) element associated with the measurement outcome. If the measurement is a projection onto a pure state, $\hat{\Pi}(\vb*{m})$ is the projection-valued measure (PVM). A Gaussian measurement has Gaussian POVM elements. Table \ref{tab:G_measure} lists common single-mode Gaussian measurements; a heterodyne detector projects onto a coherent state, a homodyne detector projects onto a rotated (unnormalizable) position eigenstate, and a generaldyne detector is a projection onto a squeezed, displaced state. Since the POVM element is a density operator
it can also be described by a Wigner function. For generaldyne multimode POVM elements,
\[
W_{\hat{\Pi}(\vb*{m})}(\vb*{q})=G_{\vb*{m},\vb*{\omega}}(\vb*{q})=\prod_{i=1}^NG_{\vb*{m}_i, \vb*{\omega}_i}(\vb*{q}_{i})
\]
 where $\vb*{m}_i$ and $\vb*{\omega}_i$ are the first and second moments of the Gaussian POVM element in each mode. 
\begin{table}[]
    \centering
    \caption{Single-mode Gaussian projectors and the associated covariance matrices $\vb*{\omega}$ and displacement vectors $\vb*{m}$ of their Wigner function. }
    \begin{ruledtabular}
    \begin{tabular}{cccc}
         Detector & POVM & $\vb*{\omega}$ & $\vb*{m}$ \\
         \colrule
         Heterodyne & $\ketbra{\alpha}$ & $\frac{\hbar}{2}\mathds{1}_2$ & $\sqrt{2\hbar}\begin{pmatrix}
             \Re\alpha \\ \Im\alpha
         \end{pmatrix}$ \\
         Homodyne ($x$) & $\ketbra{x}$ & $\lim\limits_{r\to\infty}\begin{pmatrix}
    e^{-2r} & 0 \\ 0 & e^{2r}
\end{pmatrix}$ & $\begin{pmatrix}
    x \\0 
\end{pmatrix}$ \\
Homodyne ($p$) & $\ketbra{p}$ & $\lim\limits_{r\to\infty}\begin{pmatrix}
    e^{2r} & 0 \\ 0 & e^{-2r}
\end{pmatrix}$ & $\begin{pmatrix}
    0 \\p
\end{pmatrix}$ \\
Generaldyne &$\ketbra{z,\alpha}$ & $\vb*{S}(z)\vb*{S}(z)^T$ & $\sqrt{2\hbar}\begin{pmatrix}
             \Re\alpha \\ \Im\alpha
         \end{pmatrix}$
    \end{tabular}
    \end{ruledtabular}
    \label{tab:G_measure}
\end{table}
A heterodyne detector projects mode $i$ onto a coherent state with amplitude $\alpha_i$. Homodyne detection measures an arbitrary quadrature $\hat{x}_\phi=\hat{R}(\phi)^\dagger\hat{x}\hat{R}(\phi)=\cos\phi \hat{x}+\sin\phi \hat{p}$. This can always be mapped to rotating the state by the opposite angle, and measuring the $\hat{x}$ quadrature. 
$\vb*{\mu}_{m_i}=\sqrt{2\hbar}(\Re\alpha_i,\Im\alpha_i)^T$ and $\vb*{\sigma}_{m_i}=\frac{\hbar}{2}\mathds{1}_2$, 
while for homodyne detection of the arbitrary quadrature $\hat{x}_\phi=\cos\phi \hat{x}+\sin\phi \hat{p}$ quadrature, which corresponds to rotating the state in mode $i$ by $-\phi$ and measuring $\hat{x}$, $\vb*{\mu}_{m_i}=(x_i,0)^T$ and $\vb*{\sigma}_{m_i}=\lim_{r\to\infty}\big(\begin{smallmatrix}
    e^{-2r} & 0 \\ 0 & e^{2r}
\end{smallmatrix}\big)$. 
A generaldyne measurement, a generalized case of a homodyne and heterodyne measurement, corresponds to projecting onto a squeezed, displaced state $\ket{z,\alpha}=\hat{D}(\alpha)\hat{S}(z)\ket{0}$. 

Taking the trace is equivalent to performing a phase space integral over the Wigner functions of $\hat{\rho}$ and  $\hat{\Pi}(\vb*{m})$,
\begin{equation}
    p(\vb*{m})=(2\pi\hbar)^N\int_{\mathds{R}^{2N}}\dd^{2N}{\vb*{q}} W_{\hat{\rho}}(\vb*{q})W_{\hat{\Pi}(\vb*{m})}(\vb*{q}),
\end{equation}
which for Gaussian states and Gaussian measurements has the simple Gaussian form, $p(\vb*{m})=G_{\vb*{\mu},\vb*{\sigma}+\vb*{\omega}}(\vb*{m})$. \\

If a subset of modes, labeled by $B$ is measured with POVM element $\hat{\Pi}_B(\vb*{m})=\bigotimes_{i=1}^{N_B}\hat{\Pi}_i(m_i)$, the remaining modes, labeled by $A$, will transform according to
\begin{equation}
\hat{\rho}_{A}'(\vb*{m})=\frac{\mathrm{Tr}_B[\hat{\rho}\hat{\Pi}_B(\vb*{m})]}{p(\vb*{m})},
\end{equation}
where $\Tr_B[.]$ indicates a partial trace over the modes labeled by $B$. We can partition the vector of quadratures describing $\hat{\rho}$  according to the mode labels, $\vb*{q}=(\vb*{q}_{A},\vb*{q}_{B})^T$, and arrange the covariance matrix and displacement vector in this ordering,
\begin{equation}
    \vb*{\sigma}=\begin{pmatrix}
\vb*{\sigma}_{A} & \vb*{\sigma}_{AB} \\ \vb*{\sigma}_{AB}^T & \vb*{\sigma}_{B}
\end{pmatrix}
 \quad\text{and}\quad \vb*{\mu}=\begin{pmatrix}\vb*{\mu}_{A} \\ \vb*{\mu}_{B}\end{pmatrix}.
\end{equation}

Once again, we apply the trace rule to express $\mathrm{Tr}_B[\hat{\rho}\hat{\Pi}_B(\vb*{m})]$ as a phase-space integral over the Wigner functions of $\hat{\rho}$ and $\hat{\Pi}_B(\vb*{m})$. In this case, however, the integral is taken only over the partial phase-space associated with subsystem $B$, specifically, the $\vb*{q}_B$ variables. Let $\hat{\Pi}_B(\vb*{m})$ denote a general-dyne measurement operator. When both Wigner functions are Gaussian, the result of the partial integral is another Gaussian with covariance matrix $\vb*{\sigma}_{A}'$ and displacement vector $\vb*{\mu}_{A}'(\vb*{m})$, multiplied by a Gaussian weight factor $G_{\vb*{\mu}_B,\vb*{\sigma}_B+\vb*{\omega}}(\vb*{m})$, which is the probability of the measurement result $\vb*{m}$,
\begin{align}
    \mathrm{Tr}_B[\hat{\rho}\hat{\Pi}_B(\vb*{m})]
    %&=(2\pi \hbar)^{N_{B}} \int  \dd{\vb*{q}_{B}} W_{\hat{\rho}}(\vb*{q})W_{\hat{\Pi}(\vb*{m})}(\vb*{q}_{B}) \\
    &= (2\pi \hbar)^{N_{B}}\int_{\mathds{R}^{2N_B}} \dd^{2N_B}{\vb*{q}_{B}} G_{\vb*{\sigma},\vb*{\mu}}(\vb*{q})G_{\vb*{\omega},\vb*{m}}(\vb*{q}_{B})\nonumber\\
    &=(2\pi \hbar)^{N_{B}} G_{\vb*{\mu}_B,\vb*{\sigma}_B+\vb*{\omega}}(\vb*{m}) G_{\vb*{\mu}_{A}'(\vb*{m}),\vb*{\sigma}_{A}'}(\vb*{q}_{A}),  %\\
    %p(\vb*{m})
    %&=(2\pi \hbar)^{N}\int  \dd^{2N}{\vb*{q}} W_{\hat{\rho}}(\vb*{q})W_{\hat{\Pi}(\vb*{m})}(\vb*{q}_{B}) \\
    %&=(2\pi \hbar)^N\int  \dd{\vb*{q}} G_{\vb*{\sigma},\vb*{\mu}}(\vb*{q})G_{\vb*{\sigma}_m,\vb*{\mu}_m}(\vb*{q}_{B})\\
    %&= G_{\vb*{\mu}_{B},\vb*{\sigma}_{B}+\vb*{\sigma}_{\vb*{m}}}(\vb*{m}). 
\end{align}
where
\[
&\vb*{\sigma}_{A}'=\vb*{\sigma}_{A}-\vb*{\sigma}_{AB}(\vb*{\sigma}_{B}+\vb*{\omega})^{-1}\vb*{\sigma}_{AB}^T, \label{eq:sigmaA_gauss}\\
    &\vb*{\mu}_{A}'(\vb*{m})=\vb*{\mu}_{A}+\vb*{\sigma}_{AB}(\vb*{\sigma}_{B}+\vb*{\omega})^{-1}(\vb*{m}-\vb*{\mu}_{B})\label{eq:muA_gauss}.
\]
After dividing by the probability, the Wigner function of the post-selected state in modes $A$ is another Gaussian,
\begin{equation}
    W_{\hat{\rho}_{A}'(\vb*{m})}(\vb*{q}_{A})=G_{\vb*{\mu}_{A}'(\vb*{m}),\vb*{\sigma}_{A}'}(\vb*{q}_{A}).
\end{equation}
These simple matrix transformations provide us with an elegant and fast way of working with Gaussian states, measurements and transformations.

\subsection{The Linear Combination of Gaussians Formalism}
\label{sec:sumofG}

The LCoG formalism emerges naturally for a class of non-Gaussian states. For example, multi-headed cat states are superpositions of coherent states, and approximate Gottesmann-Kitaev-Preskill (GKP) \cite{gottesman_encoding_2001} states are superpositions of squeezed, displaced states. Gaussian transformations and Gaussian measurements on such states can be conveniently represented via the Wigner function \cite{bourassa_fast_2021}. Furthermore, it can be shown that pseudo photon-number-resolving detectors also have an exact LCoG formalism, because their POVM elements can be expressed as linear combinations of thermal states, making their addition to the LCoG framework straightforward.

In this section, the Wigner function of states and POVM elements that have an exact LCoG decomposition will be described. We will review how to use the LCoG formalism to perform operations on the states, such as Gaussian gates, loss channels and measurements. Lastly, we will describe an approximate LCoG formalism for the simulation of PNRDs, which is also based on linear combinations of thermal states \cite{bourassa_fast_2021}. 

\subsubsection*{Single-mode non-Gaussian states}
We consider a pure state consisting of a superposition of general Gaussian pure states, that is, displaced, squeezed vacuum states $\ket{z, \alpha}=\hat{S}(z)\hat{D}(\alpha)\ket{0}$. We limit ourselves to the situation where each term is squeezed with the same magnitude and under the same angle,
\[
\ket{{\psi}}=\hat{S}(z)\sum_{k} a_k \ket{\alpha_k}, \quad a_k,\alpha_k, z \in \mathds{C},\label{eq:ket_sqz_gauss_superpos}
\]
where $\ket{\alpha_k}$ are coherent states with amplitude $\alpha_k$.
The density operator of this state is a linear combination of the outer product of Gaussian states,
% \[
% \ketbra{\psi} = \sum_{kl}a_k a_l^* \ketbra{\alpha_k, z_k}{\alpha_l, z_l}. \label{eq:psi_sqz_gauss_superpos}
% \]

% ($z_k=z_l=z)$. 
\[
\ketbra{\psi} = \hat{S}(z)\sum_{kl}a_k a_l^* \ketbra{\alpha_k}{\alpha_l} \hat{S}^\dagger(z) .\label{eq:dm_sqz_gauss_superpos}
\]
The Wigner-Weyl transform of the outer product of two different coherent states $\ketbra{\alpha}{\beta}$ is derived in the Appendix of \cite{bourassa_fast_2021}; it is a Gaussian with complex weights and means, and with the vacuum covariance matrix,
\[
W_{\ketbra{\alpha}{\beta}}(\vb*{q})&=e^d G_{\vb*{\mu},\vb*{\sigma}}(\vb*{q}), \label{eq:wig_weyl_coherent_outer}\quad\text{where}\quad\vb*{\sigma} = \frac{\hbar}{2}\mathds{1}_2, \\
\vb*{\mu} &= \sqrt{\frac{\hbar}{2}}\begin{pmatrix}
\Re(\alpha+\beta) +i\Im(\alpha-\beta), \label{eq:coherent_outer_mu}\\ \Im(\alpha+\beta)+i\Re(\beta-\alpha)
\end{pmatrix}, \\
 d &=-\frac{1}{2}\Im(\alpha-\beta)^2 -i\Im(\beta)\Re(\alpha) \nonumber\\
&\quad\:-\frac{1}{2}\Re(\alpha-\beta)^2+i\Im(\alpha)\Re(\beta) \label{eq:coherent_outer_coeff}.
\]
Therefore, the Wigner function of non-Gaussian states expressed in the form of Eq.\ \eqref{eq:dm_sqz_gauss_superpos} can be represented by a more general LCoG \cite{bourassa_fast_2021}, 
\begin{equation}
    W_{\hat{\rho}}(\vb*{q})=\sum_{m\in\mathcal{M}} e^{c_{m}}G_{\vb*{\mu}_{m},\vb*{\sigma}_{m}}(\vb*{q}),\label{eq:Wigner_sumofG}
\end{equation}
where $c_m$, $\vb*{\mu}_m$, and $\vb*{\sigma}_m$ are the log-coefficient, displacement vector and covariance matrix of each Gaussian in the set of indices $\mathcal{M}$. $c_m$, $\vb*{\mu}_m$, and $\vb*{\sigma}_m$ are in general allowed to be complex, as long as the state is normalized, $\sum_m e^{c_m} = 1$, and the total Wigner function is real. Furthermore, the real part of $\vb*{\sigma}_m$ should be positive definite for the state to be bounded in phase space \cite{bourassa_fast_2021}.

For states in the form of Eq.\ \eqref{eq:dm_sqz_gauss_superpos}, the covariance matrix is the same for each Gaussian, and is equal to $\vb*{\sigma}=\frac{\hbar}{2}\vb*{S}(z)\vb*{S}(z)^T$, where $\vb*{S}(z)$ is the symplectic matrix of the squeezing operation. Each displacement vector is found by squeezing the displacement vector of Eq.\ \eqref{eq:coherent_outer_mu}, $\vb*{\mu}_{kl}\mapsto \vb*{S}(z)\vb*{\mu}_{kl}$. The log-coefficient is $c_{kl} = \ln(a_k)+\ln(a_l^*) + d_{kl}$, where $d_{kl}$ can be obtained from Eq.\ \eqref{eq:coherent_outer_coeff}. For $k=l$, the coefficients and means are real, while they are complex for $k\neq l$, forming the interference fringes, i.e. the negative areas in the quasi-probability distribution function. 
The interference of the complex terms produces a real Wigner function, which can be written in the following way \cite{marshall_simulation_2023},
\[
W_{\ketbra{\psi}}(\vb*{q})=\sum_{k}e^{c_{kk}} G_{\vb*{\mu}_{kk},\vb*{\sigma}}(\vb*{q}) + 2\Re\left[\sum_{l>k} e^{c_{kl}} G_{\vb*{\mu}_{kl},\vb*{\sigma}}(\vb*{q})\right],\label{eq:wigner_sum_of_G_reduced}
\]
with the norm $\mathcal{N}=\sum_k e^{c_{kk}}+2\Re[\sum_{l>k}e^{c_{l}}]$. Expressing the state in this form reduces the number of complex coefficients and means that need to be tracked by approximately a factor of two. This simplification is specific to states of the form given in Eq.\ \eqref{eq:dm_sqz_gauss_superpos}, which do not necessarily have to be pure - the $a_ka_l^*$ coefficient in Eq.\ \eqref{eq:dm_sqz_gauss_superpos} can be arbitrary. We refer to the representation in Eq.\ \eqref{eq:wigner_sum_of_G_reduced} as the reduced linear combination of Gaussians (red-LCoG) framework.

Here, we emphasize the use of the log of the coefficients because $d$ can become highly negative for, e.g. squeezed cat states with very large amplitudes, resulting in extremely small coefficients. Conversely, the coefficients can also become very large when representing pseudo photon-number-resolving detectors due to the binomial coefficients. Tracking the log-coefficient, instead of the actual coefficient is a strategy that can be used to suppress rounding errors when working with finite floating point precision.

\subsubsection*{Multimode non-Gaussian states}
The Wigner function of a class of $N$-mode non-Gaussian states can also be expressed in the form of Eq.\ \eqref{eq:Wigner_sumofG}, where $\vb*{q}\in\mathds{R}^{2N}$ is the multi-mode phase-space coordinate, $\vb*{\mu}_m\in\mathds{C}^{2N}$, and $\vb*{\sigma}_m=\mathds{C}^{2N}\times\mathds{C}^{2N}$. The tensor product of two states of the form of Eq.\ \eqref{eq:Wigner_sumofG} can be computed directly by multiplying the their Wigner functions,
\[
W_{\hat{\rho}_{1}\otimes\hat{\rho}_{2}}(\vb*{q})=\sum_{m\in\mathcal{M}}\sum_{n\in\mathcal{N}}e^{c_{m}+c_{n}}G_{\vb*{\mu}_{m}\oplus \vb*{\mu}_{n},\vb*{\sigma}_{m}\oplus \vb*{\sigma}_{n}}(\vb*{q}),
\]
which results in a multiplicative increase in the number of weights, means and covariance matrices required to represent the joint state. 

\subsubsection*{Gaussian channels}
Gaussian channels, which include symplectic operations with $\vb*{X}=\vb*{S}$ and $\vb*{Y}=\vb*{0}$, will transform each covariance matrix and displacement vector according to the map; $\vb*{\mu}_{m}\mapsto \vb*{X}\vb*{\mu}_{m}+\vb*{d}$ and $\vb*{\sigma}_{m}\mapsto \vb*{X}\vb*{\sigma}_{m}\vb*{X}+\vb*{Y}$, making them extremely simple to implement in this formalism. 

\subsubsection*{Measurements}
The update rules due to partial Gaussian measurements in Section \ref{sec:gauss_formalism} can easily be extended to states and POVM elements in the form of Eq.\ \eqref{eq:Wigner_sumofG}. In the following, we describe the update rules for partial measurements on multimode states where both the state and the POVM element are written as a LCoG.

As before, the state $\hat{\rho}$ is divided into two sets of modes; we measure the modes labeled by $B$, and compute the heralded state in the modes labeled by $A$. First, we partition each covariance matrix and displacement vector into these sets,
\begin{equation}
    \vb*{\sigma}_m=\begin{pmatrix}
\vb*{\sigma}_{m,A} & \vb*{\sigma}_{m,AB} \\ \vb*{\sigma}_{m,AB}^T & \vb*{\sigma}_{m,B}
\end{pmatrix}
 \quad\text{and}\quad \vb*{\mu}_m=\begin{pmatrix}\vb*{\mu}_{m,A} \\ \vb*{\mu}_{m,B}\end{pmatrix}.
\end{equation}
The modes in $B$ are projected onto a POVM element, $\hat{\Pi}_{B}(\vb*{m})$, whose Wigner function can be written as, 
\[
W_{\hat{\Pi}_{B}(\vb*{m})}=\sum_{n}e^{d_{n}}G_{\vb*{\nu}_{n},\vb*{\omega}_{n}}(\vb*{q}_{B}), \label{eq:POVM_sumofG}
\] 
where $d_n\in\mathds{C}$, $\vb*{\nu}_n\in\mathds{C}^{2N_B}$, and $\vb*{\omega}_n\in\mathds{C}^{2N_B}\times \mathds{C}^{2N_B}$ are the log-coefficients, means, and covariance matrices of each Gaussian in the POVM element. The partial measurement can be performed by considering the transformation of each product of Gaussians,  

\begin{align}
    &\mathrm{Tr}[\hat{\rho}\hat{\Pi}_{B}(\vb*{m})]=\nonumber\\
    &(2\pi\hbar)^{N_B} \int  \dd^{2N_B}{\vb*{q}_{B}}\left[\sum_{m}e^{c_{m}}G_{\vb*{\sigma}_{m},\vb*{\mu}_{m}}(\vb*{q})\sum_{n}e^{d_{n}}G_{\vb*{\nu}_{n},\vb*{\omega}_{n}}(\vb*{q}_{B}) \right]\nonumber\\
    &=\sum_{mn}e^{c_{n}+d_{m}+\gamma_{mn}}G_{\vb*{\sigma}_{mn,A}',\vb*{\mu}_{mn,A}'}(\vb*{q}_{A}),
\end{align}
where
\begin{align}
    \vb*{\sigma}_{mn,A}'&=\vb*{\sigma}_{m,A}-\vb*{\sigma}_{m,AB}(\vb*{\sigma}_{m,B}+\vb*{\omega}_{n})^{-1}\vb*{\sigma}_{m,AB}^T \label{eq:sigmaA_sum_of_G},\\
    \vb*{\mu}_{mn,A}'&=\vb*{\mu}_{m,A}+\vb*{\sigma}_{m,AB}(\vb*{\sigma}_{m,B}+\vb*{\omega}_{n})^{-1}(\vb*{\nu}_{n}-\vb*{\mu}_{m,B})\label{eq:muA_sum_of_G},\\
    e^{\gamma_{mn}}&=(2\pi\hbar)^{N_B}G_{\vb*{\sigma}_{m,B}+\vb*{\omega}_{n},\vb*{\mu}_{m,B}}(\vb*{\nu}_{n})\label{eq:gamma_sum_of_G}.
\end{align}
The probability of the measurement outcome is simply a summation over the new weights,
\begin{equation}
    p(\vb*{m})=\sum_{mn}e^{c_{m}+d_{n}+\gamma_{mn}}.
\end{equation}
The log-weight factor $\gamma_{mn}$ is in general different for each $mn$, and thus will not cancel when dividing the output state by the measurement probability. 
The post-selected state in can be written as
\[
W_{\hat{\rho}_{A}'(\vb*{m})}(\vb*{q}_{A})=\frac{1}{p({\vb*{m})}}\sum_{mn} e^{c_n+d_m+\gamma_{mn}}G_{\vb*{\mu}_{mn,A}',\vb*{\sigma}_{mn,A}'}(\vb*{q}_{A}).
\]
Thus, when non-Gaussian measurements are performed on non-Gaussian states of the form of Eq.\ \eqref{eq:Wigner_sumofG}, the number of Gaussians increases multiplicatively and each weight is multiplied by a Gaussian factor. 

In the remainder of this section, we turn our attention to photon-counting POVMs that can be written as a linear combination of Gaussians as in Eq.\ \eqref{eq:POVM_sumofG} and implemented in the manner described in the following.

\subsubsection*{On/off detectors}
A threshold (on/off) detector, also known as a click detector, which can only distinguish between the presence or absence of photons, is a simple example of a pseudo-photon-number-resolving POVM that admits a LCoG description. Its POVM elements are the vacuum projector for no clicks, $\hat{\Pi}_0$, and the complement of the vacuum projector for a click, $\hat{\Pi}_1$,
\[
\hat{\Pi}_0 =\ketbra{0}, \qand \hat{\Pi}_{1}=\hat{\mathds{I}}-\ketbra{0}.
\]
The Wigner function of the POVM elements are
\[
W_{\ketbra{0}}(\vb*{q})&=G_{\vb*{0}, \frac{\hbar}{2}\mathds{1}}(\vb*{q}), \\
W_{\hat{\mathds{I}}-\ketbra{0}}(\vb*{q})& = (2\pi\hbar)^{-1} - G_{\vb*{0}, \frac{\hbar}{2}\mathds{1}}(\vb*{q}),\label{eq:wig_click}
\]
where the Wigner function of the identity operator is a constant. Click detectors fit perfectly within the LCoG framework because the no-click detection is a projection onto a Gaussian (a heterodyne measurement), and the click detection is a linear combination of two Gaussians.  For example, if we detect a click in mode $j$, the output state is proportional to
\[
W_{\hat{\rho}'}(\vb*{q})\propto\int \dd^2{\vb*{q}_{j}} W_{\hat{\rho}}(\vb*{q})-2\pi\hbar\int \dd^2{\vb*{q}_{j}} W_{\hat{\rho}}(\vb*{q})W_{\ketbra{0}}(\vb*{q}_{j}).
\]
If the initial state $\hat{\rho}$ is a LCoG, the first term is a trace over the $j$'th mode, which corresponds to simply removing the elements associated with that mode from the covariance matrix and displacement vector. The second term is the partial overlap with vacuum in mode $j$, giving
\[
W_{\hat{\rho}'}(\vb*{q})=\frac{1}{1-p_0}\sum_{k\in\mathcal{K}} \left[c_k G_{\vb*{\mu}_{k,A},\vb*{\sigma}_{k,A}}(\vb*{q}_A)-\gamma_k G_{\vb*{\mu}_{k,A}',\vb*{\sigma}_{k,A}'}(\vb*{q}_A)\right],
\]
where $\gamma_k =2\pi\hbar c_k G_{\vb*{\mu}_{k,B},\vb*{\sigma}_{k,B}+\vb*{\sigma}_j}(\vb*{\mu}_j)$, $\vb*{\sigma}_j=\frac{\hbar}{2}\mathds{1}_2$, and $\vb*{\mu}_j=\vb{0}$. $p_0=\Tr[\hat{\rho}\ketbra{0}_j]$ is the probability of measuring vacuum in mode $j$.
% When performing an "on" measurement, one traces over the $B$ modes in the first term, and projects onto vacuum in the second term, according to the Gaussian transformation rules in Eqs. \eqref{eq:sigmaA_sum_of_G}-\eqref{eq:weights_sum_of_G}

\subsubsection*{Pseudo photon-number-resolving detectors}

%When approximating
% 

A pseudo photon-number detector is a fan-out onto $M$ on/off detectors. For pseudo-PNRD, the POVM element for $k$ clicks can be expressed as a linear combination of $k+1$ thermal states with different average photon numbers \cite{bressanini_gaussian_2024},
\begin{equation}
    \hat{\Pi}_{k,M} = \mqty(M\\ k) \sum_{l=0}^k \mqty(k \\ l)(-1)^l \frac{1}{\eta_{kl} }\hat{\rho}_\text{th}\left(\bar{n}_{kl}\right)\label{eq:pPNRD_POVM},
\end{equation}
where $k\in\{0,\dots,M\}$ is the number of clicks, $\eta_{kl} = (M-k+l)/M$ and $\bar{n}_{kl}=(1-\eta_{kl})/\eta_{kl}$ is the occupation number of each thermal state. The derivation of Eq.\ \eqref{eq:pPNRD_POVM} is given in App. \ref{app:ppnrd}. The Wigner function of the POVM is a sum of $k+1$ Gaussians with the following coefficients, means and covariance matrices in Eq.\ \eqref{eq:Wigner_sumofG},
\[
e^{c_{l}}&=\begin{pmatrix}
M \\ k
\end{pmatrix}\begin{pmatrix}
k \\ l
\end{pmatrix}
\frac{(-1)^{l}}{\eta_{kl}}, \\
\vb*{\mu}_{l}&=(0,0)^{T},\\
\vb*{\sigma}_{l}&=\frac{\hbar}{2}(2\bar{n}_{kl}+1)\mathds{1}_{2}.
\]

Note that in the limit of $\eta_l=0$, we recover the identity operator as  $\eta_l^{-1} \hat{\rho}_\text{th}(\bar{n}=\infty)=\sum_{n=0}^\infty \ketbra{n} = \hat{\mathds{I}}$. Conversely, for $\eta_l = 1$, $\bar{n}=0$, we recover the vacuum state. The case $M=1$ corresponds to the single on/off detector POVM from the previous section.

\subsubsection*{Photon number resolving detectors}
\begin{figure}
    \centering
\includegraphics[width=\linewidth]
{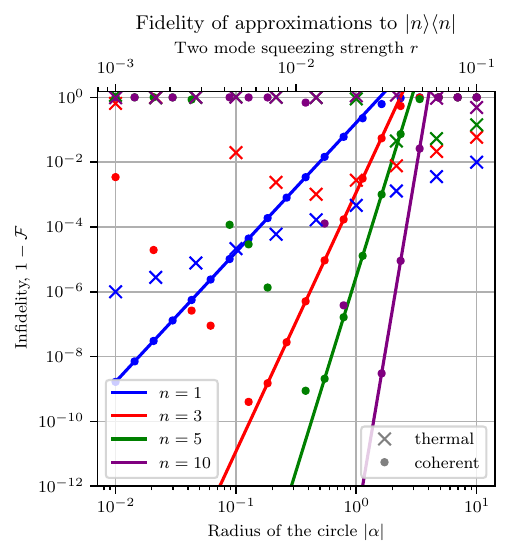}
    \caption{\textbf{The numerical fidelity of Fock approximations with LCoGs}. The fidelity of the analytical $\ketbra{n}$ with i) the thermal state approximation in Eq.\ \eqref{eq:Fock_thermal} with squeezing strength $r$ (crosses), ii) the coherent state decomposition in Eq.\ \eqref{eq:wigner_fock_approx}  with circle radius $\abs{\alpha}$ (dots). The solid line is the theoretical fidelity of the coherent state decomposition. The overlaps are computed numerically by integrating over the product of the Wigner functions on a fine grid of phase space points.}
    \label{fig:fock_approx}
\end{figure}

So, far we have considered pseudo-PNRDs, which can only approximately resolve photons, in the LCoG framework. True PNRDs have a photon number basis description, for which an exact LCoG decomposition does not exist. In order to merge this with the LCoG framework, we must introduce some approximations. One possible way to approximate a Fock state with a LCoG is derived in Bourassa et al. \cite{bourassa_fast_2021}; the $n$'th Fock state can be written as a linear combination of $n+1$ thermal states with different temperatures,
\[
\ketbra{n} \approx \frac{1}{\mathcal{N}_n}\sum_{k=0}^{n} c_k \hat{\rho}_{\text{th}, k} .\label{eq:Fock_thermal}
\]
The Wigner function of the approximation is of the form Eq.\ \eqref{eq:Wigner_sumofG}, with coefficients\footnote{The parity of the coefficient has a minor correction here.}, covariances and means given by 
\[
&c_k = (-1)^{k}\mqty(n\\k)\left[\frac{1-nr^2}{1-(n-k)r^2}\right], \label{eq:fock_th_coeff}\\
&\vb*{\sigma}_k=\frac{\hbar}{2}\frac{1+(n-k)r^2}{1-(n-k)r^2}\mathds{1}_2,\label{eq:fock_th_cov}\\
&\vb*{\mu}_k=\vb{0}\label{eq:fock_th_mean}.
\]
The parameter $r$ arises from the derivation of the approximation, which is based on the concept of heralded photon addition. Specifically, two-mode squeezing operations with strength $r$, ancillary vacuum modes, and click detectors are used to approximate the $\hat{a}^{\dagger n}$ operation on vacuum. The normalization constant $\mathcal{N}_n$, corresponds to the success probability of the photon addition protocol, for which all on/off click detectors register a click. In order to minimize the probability of the click detectors receiving more than one photon, which decreases the fidelity of the output state to Fock $n$, the squeezing must be kept low. However, this in turn reduces the probability of any click occurring, thereby lowering $\mathcal{N}_n$. For $r<n^{-\frac{1}{2}}$, the approximation returns a physical state, while $r\to 0$ should perfectly approximate the Fock state. The numerical fidelity of the approximation is plotted in Fig.\ \ref{fig:fock_approx}.

The Fock approximation is similar to the pPNRD POVM elements in Eq.\ \eqref{eq:pPNRD_POVM}, which is also a linear combination of thermal states. A pPNRD is an approximation of a PNRD, and a straightforward way to improve the approximation is to increase the number of on/off detectors $M$ in the fan-out. In fact by increasing $M$ to several orders of magnitude, one can recover similar coefficients, means and covariances as in Eq.\ \eqref{eq:fock_th_coeff}, \eqref{eq:fock_th_cov}, \eqref{eq:fock_th_mean}. 

In practice, using the Fock state approximation in Eq.\ \ref{eq:Fock_thermal} or increasing $M$ in Eq.\ \eqref{eq:pPNRD_POVM} to approximate PNRDs in the LCoG formalism proves to be unsuitable for practical simulations. This is due to the poor numerical stability because of the low value of $\mathcal{N}_n$, and correspondingly large binomial coefficients. Even when tracking the log-weights, we encounter rounding errors due to finite floating point precision when attempting to calculate the norm. Therefore, other Fock state approximations must be considered in order to add PNR capabilities to the LCoG framework.

\subsection{The coherent state decomposition}
\label{sec:coherent_state_decomp}
A different way of representing Fock states as a LCoG was recently introduced by Marshall and Anand \cite{marshall_simulation_2023}, in which photon number states are approximated by a superposition of a finite number of coherent states, characterized by a finite coherent rank. 
In \cite{marshall_simulation_2023}, the framework focuses on the simulation of linear optical circuits with pure, non-Gaussian input states. 
% In this decomposition, the squeezing operation is not "free" - it actively increases the number coherent states and thereby the simulation complexity. 
In the following, we will calculate the Wigner function of this Fock approximation and discuss its suitability for the LCoG framework. One benefit of working with the Wigner function is that active Gaussian operations such as squeezing and displacement can be performed without increasing the simulation complexity, thereby capturing the minimal non-Gaussian resources that are necessary to represent mixed states. 

% Inspired by finite rank stabilizer decompositions \cite{bravyi_simulation_2019} for quantum circuit simulation,   

\begin{figure}
    \centering
    \includegraphics[width=\linewidth]{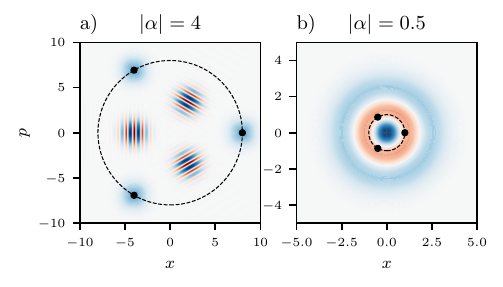}
    \caption{\textbf{Coherent state decomposition of a two-photon Fock state}. The Wigner function of the approximation of $\ketbra{2}$ in Eq.\ \eqref{eq:wigner_fock_approx} for two different coherent state amplitudes with magnitude (radius) $\abs{\alpha}$. The center of each of the three coherent states in the superposition is plotted by the blue circle along the circumference. The center of the complex Gaussians that form the interference fringes lie inside the circle. }
    \label{fig:fock_coherent_radius}
\end{figure}

\subsubsection*{Fock state}\label{sec:fock_approx_coherent}
The $n$'th photon number state can be approximated by a superposition of $n+1$ uniformly distributed coherent states on a small ring in phase space \cite{marshall_simulation_2023},
\begin{equation}
% \left|n\right> \approx \frac{1}{\sqrt{\mathcal{N}}}\frac{\sqrt{n!}}{n+1}\frac{e^{\epsilon^2/2}}{\epsilon^n}\sum_{k=0}^n c_k \left|\alpha_k\right>,\label{eq:fock_coherent_ket}
\left|n\right> \approx \frac{1}{\sqrt{\mathcal{N}}}\sum_{k=0}^n c_k \left|\alpha_k\right>,\label{eq:fock_coherent_ket}
\end{equation}
where the displacement of the coherent states is $\alpha_{k}=\epsilon e^{2\pi ik/(n+1)}$, $\epsilon$ is the radius of the ring, and the coefficient is $c_{k}= e^{-2\pi ikn/(n+1)}$. $\epsilon$ is a free parameter and determines the fidelity of the approximation, which is on the order of $\mathcal{F}=1-\mathcal{O}\left(\frac{n!}{(2n+1)!}\epsilon^{2(n+1)}\right)$, and is directly related to the norm $\mathcal{F}=\mathcal{N}^{-1}$, so a smaller radius results in a better approximation. However, a photon number state with high $n$ can be well-approximated with a relatively large radius, as shown in Fig.\ \ref{fig:fock_approx}, compared to smaller $n$.\\

The density operator of the approximate photon number state in Eq.\ \eqref{eq:fock_coherent_ket} is a sum of $(n+1)^2$ coherent state outer products,
\[
\ketbra{n}\approx \frac{1}{\mathcal{N}}\sum_{k,l=0}^n c_{k}c_{l}^*\ketbra{\alpha_{k}}{\alpha_{l}}. \label{eq:fock_coherent_dm}
\]
Thus, the LCoG representation for Fock state $n$ in the coherent state decomposition uses $(n+1)^2$ Gaussians. The Wigner function is,
\[
W_{\ketbra{n}}(\vb*{q}) = \sum_{k,l=0}^n e^{w_{kl}} G_{\vb*{\mu}_{kl},\frac{\hbar}{2}\mathds{1}_2}(\vb*{q}),\label{eq:wigner_fock_approx}
\]
where each log-weight is,
\[
w_{kl} = d_{kl} -\frac{2\pi i(k-l)n}{n+1}\in\mathds{C}, \label{eq:weights_fock_povm}
\]
where $d_{kl}$ can be evaluated using Eq.\ \eqref{eq:coherent_outer_coeff}. Each displacement vector $\vb*{\mu}_{kl}$ can be evaluated with Eq.\ \eqref{eq:coherent_outer_mu}.
% \[
% \vb*{\mu}_{kl}&=\sqrt{\frac{\hbar}{2}}\begin{pmatrix}
% \mathrm{Re}(\alpha_{k}+\alpha_{l})+i\mathrm{Im}(\alpha_{k}-\alpha_{l}) \\
% \mathrm{Im}(\alpha_{k}+\alpha_{l})+i\mathrm{Re}(\alpha_{l}-\alpha_{k})
% \end{pmatrix}\in\mathds{C}^2.
% \]
The state can be normalized by dividing each weight with the sum of the weights, $\mathcal{N}=\sum_{k,l=0}^n e^{w_{kl}}$.

In Fig.\ \ref{fig:fock_approx}, we compare the fidelity and numerical stability of the two Fock state approximations in the LCoG formalism. The fidelity of the approximation using coherent states in Eq.\ \eqref{eq:wigner_fock_approx} is several orders of magnitude better than the fidelity of the thermal state approximation in Eq.\ \eqref{eq:Fock_thermal}. At the same time, the coherent state decomposition is numerically stable at higher photon numbers. This is because there are no binomial coefficients, and the weights in Eq.\ \eqref{eq:weights_fock_povm} are in general small. For each $n$, there should be an $\abs{\alpha}$ interval for which it is possible to have a numerically stable and high fidelity approximation. This allows us to add PNRD with relatively high photon numbers to the LCoG framework. 
\subsubsection*{Connection to the Fock basis}
The coherent state decomposition can be extended to approximate arbitrary superpositions of photon number states \cite{marshall_simulation_2023}. Let $\ket{\psi}=\sum_{l=0}^n a_{l}\ket{l}$ be a superposition of photon number states up to photon number $n$ with coefficients $a_l\in\mathds{C}$. $\ket{\psi}$ can be approximated by
%(see Eq.\ (22) of \cite{marshall_simulation_2023}). 
\begin{equation}
    \ket{\psi}\approx \frac{1}{\sqrt{\mathcal{N}}}\sum_{k=0}^n c_k \ket{\alpha_k},\label{eq:fock_superpos_coherent}
\end{equation}
where $\alpha_k = \epsilon e^{2\pi i k /(n+1)}$ and the coefficients are
\begin{align}
    c_k &= \sum_{l=0}^n \sqrt{l!}\frac{a_l} {\epsilon^l}e^{-2\pi i lk /(n+1)} \label{eq:fock_superpos_coeffs}.  
\end{align} 
The fidelity of the approximation is again determined by the radius $\epsilon$,
$\mathcal{F}=1-\mathcal{O}\left(\frac{1}{(n+1)!}\epsilon^{2(n+1)}\right)=\mathcal{N}^{-1}$, which must be smaller than the radius of the approximation of a single Fock state in Eq.\ \eqref{eq:fock_coherent_ket} to achieve the same fidelity. The Wigner function of this approximation has the same form as Eq.\ \eqref{eq:wigner_fock_approx}, just with different weights. This means that we can simulate any non-Gaussian state of stellar rank $n$ in the LCoG framework.

Additionally, any truncated operator in the Fock basis can also be expressed as a linear combination of coherent state outer products (see Eq.\ (36) in \cite{marshall_simulation_2023}). This establishes a direct connection between Fock-basis simulations and the LCoG framework, allowing conversion between the two representations. Any density operator in the form of Eq.\ \eqref{eq:dm_sqz_gauss_superpos}, where the squeezing operator is replaced by a general Gaussian operator, can be easily converted into the Fock basis, c.f. App. \ref{app:coherent_to_fock}.

\subsubsection*{Suitability for the LCoG framework}
The main advantage of using the Wigner function of the coherent state formalism is that it eliminates the need to specify a photon number cutoff.  Active operations, i.e. squeezing and displacement, can be applied on a non-Gaussian state without increasing the number of weights, since the operation only changes the means and covariance matrix of the Wigner function. The framework is fast; it involves only a single covariance matrix, and the transformations operate with matrices that have relatively small dimensions (matrices of size $2N\times 2N$, where $N$ is the number of modes), compared to the dimension of the tensor in Fock basis simulations. Additionally, the operations on each mean and weight can be parallelized. 

The challenge of using the coherent state decomposition in the LCoG framework is that, while the infidelity of the approximation can be arbitrarily small, in practice the achievable fidelity is limited by the numerical stability (albeit to a much lesser extent than the Fock approximation using thermal states). This is discussed in more detail in Sec.\ \ref{sec:numerical_stability}. The complexity of simulating  Gaussian operations on multi-mode states with stellar rank $n$ \cite{chabaud_stellar_2020} scales exponentially, requiring $\mathcal{O}( (n+1)^{2N})$ weights and displacement vectors, similar to the number of elements in the Fock density operator with cutoff $n$. However, unlike the Fock representation, the LCoG framework does not require an increased cutoff when applying active Gaussian operations.

We can also reduce the number of Gaussians in Eq.\ \eqref{eq:wigner_fock_approx} by around half from $(n+1)^2$ to $n(n+1)/2$ by considering the real parts of complex interference fringes in the red-LCoG in Eq.\ \eqref{eq:wigner_sum_of_G_reduced}. When using this reduced representation, the complex terms must be treated carefully, especially during measurements or when calculating overlaps. These special strategies are described in more detail in App. \ref{app:reduced_formalism}. 
\section{Simulation framework}
\label{sec:lcog_simulation}
\begin{figure*}
    \centering
    \includegraphics[width=0.75\linewidth]
    {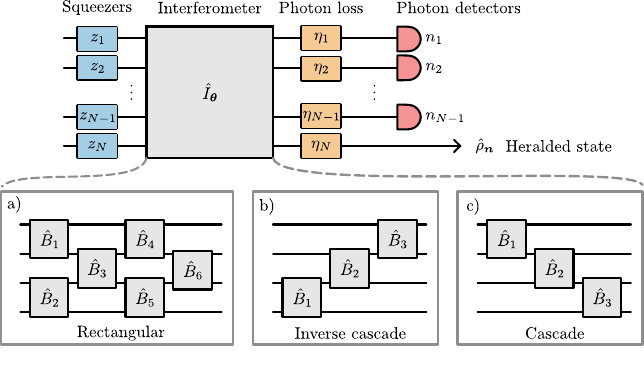}
    \caption{\textbf{An $N$-mode Gaussian Boson sampling circuit for heralded state preparation}. Vacuum is sent through $N$ single-mode squeezers with squeezing parameter $z_j=r_je^{i\phi_j}$. Next, the squeezed vacuum states are sent through an interferometer with a certain beam splitter topology. Each beam splitter has a transmissivity angle $\theta_k$ and phase $\phi_k$. If specified, the photon loss channel with transmissivity $\eta_j$ is applied on  each mode. Lastly, all but the last mode is measured with PNRDs, heralding the output state. a)-c) The three different beam splitter topologies are sketched for $N=4$ modes. The rectangular/Clements interferometer offers full connectivity with $N(N-1)/2$ beam splitters, while the cascades offer partial connectivity, but use only $N-1$ beam splitters.   }
    \label{fig:GBS_circuits}
\end{figure*}
In this section, we describe how the ideas from the coherent state decomposition are incorporated into the LCoG framework for the simulation and optimization of CV quantum optical circuits. We use the example of simulating non-Gaussian state preparation with a Gaussian Boson sampling device to showcase the utility of the framework. First, we show how to simulate the heralded output state with the partial measurement rules from Sec.\ \ref{sec:sumofG} using the PNRD approximation from Sec.\ \ref{sec:coherent_state_decomp}. Then, we show that evaluating quality measures of the output state, such as the quantum fidelity and two figures of merit related to GKP error correction properties: the effective squeezing \cite{duivenvoorden_single-mode_2017} and the GKP nonlinear squeezing \cite{marek_ground_2024}, can be done straightforwardly in the LCoG framework. These quality measures can be used as cost functions in an optimizer and we derive the analytical gradients of these figures of merit with respect to the Gaussian circuit parameters. Lastly, we outline a strategy to reduce the number of Gaussians needed to represent the heralded state, which can help reduce the memory requirements of further operations on the state, such as breeding \cite{weigand_generating_2018, takase_generation_2024, aghaee_rad_scaling_2025}.

\subsection{Simulation overview}
\label{sec:output_state}
The core of the LCoG framework revolves around keeping track of the log-coefficients $w_k$, mean vectors $\vb*{\mu}_k$ and covariance matrices $\vb*{\sigma}_k$ of states in the from of Eq.\ \eqref{eq:Wigner_sumofG} or Eq.\ \eqref{eq:wigner_sum_of_G_reduced}. For example, the state $\hat{\rho}$ can be represented as the \texttt{State} object in Python with attributes \texttt{num\_modes}, \texttt{log\_weights}, \texttt{means}, and \texttt{covs}, as listed in Table \ref{tab:state_attributes}. Various operations, such as Gaussian channels and measurements as listed in Table \ref{tab:state_operations}, can be performed on the state by updating the state attributes.

\begin{table}[]
\centering
\caption{Attributes of the \texttt{State} object representing a multi-mode state in LCoG formalism in Eq.\ \eqref{eq:Wigner_sumofG} and Eq.\ \eqref{eq:wigner_sum_of_G_reduced}.}
\begin{ruledtabular}

\begin{tabular}{lp{45mm}p{20mm}}
%\hline\hline
Attribute & Description & Dimension \\ 
%\hline
\colrule
\texttt{num\_modes} & No. of modes, $N$ &  1\\
\texttt{num\_weights} & No. of weights, $n_w$ & 1\\
\texttt{log\_weights} & List of log-weights & $(n_w,1)$ \\
\texttt{means} & List of displacement vectors & $(n_w,2 N)$ \\
\texttt{covs} & List of covariance matrices & $(n_w,2 N,2N)$  or $(1,2 N,2N)$ \\
\texttt{norm} & Sum of the weights & 1 \\
\texttt{num\_k} & No. of terms in the first sum of the red-LCoG in Eq.\ \eqref{eq:wigner_sum_of_G_reduced}. If using the full representation in Eq.\ \eqref{eq:Wigner_sumofG}, it is equal to $n_w$  & 1
%\\ \hline\hline
\end{tabular}
\end{ruledtabular}
\label{tab:state_attributes}
\end{table}

\begin{table}[]
\caption{Operations on the \texttt{State} object.}
\begin{ruledtabular}
\centering
\begin{tabular}{lp{45mm}}
%\hline\hline
Method & Description  \\
\hline
\texttt{apply\_symplectic} & Apply a Gaussian operation \\
\texttt{apply\_loss} & Apply the photon loss channel \\
\texttt{apply\_gain} & Apply the gain channel \\
\texttt{add\_state} & Tensor product with another \texttt{State} in the form of Eq.\ \eqref{eq:Wigner_sumofG} or \eqref{eq:wigner_sum_of_G_reduced} \\
\texttt{post\_select\_fock\_coherent} & Post-select with a PNRD in the coherent state decomposition in Eq.\ \eqref{eq:fock_coherent_dm} \\
\texttt{post\_select\_fock\_thermal} & Post-select with a PNRD in the thermal state decomposition in Eq.\ \eqref{eq:Fock_thermal}   \\
\texttt{post\_select\_ppnrd\_thermal} & Post-select with a pseudo-PNRD in Eq.\ \eqref{eq:pPNRD_POVM}   \\
\texttt{post\_select\_homodyne} &  Post-select on a homodyne measurement, see App. \ref{app:homodyne} \\
\texttt{get\_wigner} & Compute the Wigner function on a phase space grid (for plotting)
%\\ \hline\hline
\end{tabular}
\end{ruledtabular}
\label{tab:state_operations}
\end{table}

\subsubsection*{State preparation with a Gaussian Boson sampler}
The following code snippet corresponds to the simulation of the GBS circuit in Fig.\ \ref{fig:GBS_circuits}.
\begin{lstlisting}[language=Python]
def herald_with_GBS(N, thetas, etas, photon_pattern):
    state = State(N) #N-mode vacuum
    #build the N-mode symplectic from the circuit parameters
    S = get_symplectic(thetas) 
    state.apply_symplectic(thetas)
    state.apply_loss(etas)
    mode = 0 #always measure the first mode
    for n in photon_pattern:
        #post select on photon number n 
        state.post_select_fock_coherent(mode, n)
    
\end{lstlisting}
First, we initialize an $N$-mode vacuum state with a single unit weight $e^{w_1}=1$, a null $N$-mode displacement vector $\vb*{\mu}_1=\vb*{0}_{2N}$, and an $N$-mode vacuum covariance matrix $\vb*{\sigma}_1=\frac{\hbar}{2}\mathds{1}_{2N}$. We apply the Gaussian circuit parameterized by $\vb*{\theta}$ and any losses $\vb*{\eta}$ by applying Gaussian transformations as discussed in Sec.\ \ref{sec:sumofG}. We assume zero displacements, but they can be added at any point in the simulation, and are relevant for the preparation of asymmetrical non-Gaussian states, such as the cubic phase state \cite{gottesman_encoding_2001}. Now, the displacement vector and covariance  matrix are
given by,
\[
\vb*{\mu}_1&=\vb*{X}(\vb*{\eta})\vb*{0}_{2N}, \nonumber\\
\vb*{\sigma}_1&=\frac{\hbar}{2}\vb*{X}(\vb*{\eta})\vb*{S}(\vb*{\theta})\vb*{S}(\vb*{\theta})^T\vb*{X}(\vb*{\eta})^T+\vb*{Y}(\vb*{\eta}), \label{eq:init_gauss_state}
\]
Our task is to compute the single-mode output state of the GBS circuit post-selected on a photon pattern $\vb*{n}=(n_1,\dots,n_{N-1})^T$. Assuming perfect PNRDs, we perform the projection mode-by-mode, starting with the first mode. The projection of the first mode onto $\ketbra{n_1}$, which is approximated by $(n_1+1)^2$ Gaussians with index set $\mathcal{J}_1$ in Eq.\ \eqref{eq:wigner_fock_approx}, results in a transformation of the covariance matrix, means, and weights according to Eq.\ \eqref{eq:sigmaA_sum_of_G}, \eqref{eq:muA_sum_of_G}, and \eqref{eq:gamma_sum_of_G}.
% \[
% \vb*{\sigma}=\frac{\hbar}{2}\vb*{S}\vb*{S}^T=\begin{pmatrix}
% \vb*{\sigma}_{A} & \vb*{\sigma}_{AB} \\
% \vb*{\sigma}_{BA} & \vb*{\sigma}_{B} 
% \end{pmatrix},
% \quad \vb*{\mu}=\vb*{d}=\begin{pmatrix}
% \vb*{\mu}_{A} \\
% \vb*{\mu}_{B}
% \end{pmatrix}
% \]
We label the indices of the first mode with $B$, and the indices of the remaining modes with $A$. The (unnormalized) conditional output state is
\[
W_{\hat{\rho}_A(n_1)}(\vb*{q}_A) &\propto \sum_{j\in\mathcal{J}_1}e^{w_j}G_{\vb*{\nu}_{j},\vb*{\sigma}_{B}+\frac{\hbar}{2}\mathds{1}_2}(\vb*{\mu}_{B})G_{\vb*{\tilde{\mu}}_{j},\vb*{\tilde{\sigma}}}(\vb*{q}_A),\nonumber\\
&=\sum_{j\in\mathcal{J}_1}e^{\tilde{w}_j}
G_{\vb*{\tilde{\mu}}_{j},\vb*{\tilde{\sigma}}}(\vb*{q}_A),\label{eq:post_select_fock_Wigner}
\]
where $\tilde{w}_j=w_j+\gamma_j$, and 
\[
\gamma_{j}=-\frac{1}{2}(\vb*{\nu}_{j}-\vb*{\mu}_{B})^T(\vb*{\sigma}_{B}+\frac{\hbar}{2}\mathds{1}_2)^{-1}(\vb*{\nu}_{j}-\vb*{\mu}_B) +\delta \label{eq:gamma_GBS},\\
\tilde{\vb*{\mu}}_{j}=\vb*{\mu}_{A}+\vb*{\sigma}_{AB}(\vb*{\sigma}_{B}+\frac{\hbar}{2}\mathds{1}_2)^{-1}(\vb*{\nu}_{j}-\vb*{\mu}_{B})\label{eq:mu_GBS},\\
\tilde{\vb*{\sigma}}=\vb*{\sigma}_{A}-\vb*{\sigma}_{AB}(\vb*{\sigma}_{B}+\frac{\hbar}{2}\mathds{1}_2)^{-1}\vb*{\sigma}_{BA}\label{eq:sigma_GBS}.
\]
% $\vb*{\nu}_{j}=\bigoplus_{i=1}^{N-1}\vb*{\nu}_{i}^{[j]}$
$w_j$ and $\vb*{\nu}_{j}$ are the log-weight and displacement vector of the $j$'th Gaussian in the expression for $\ketbra{n_1}$, and $\delta = -\frac{1}{2}\ln(2\pi\det(\vb*{\sigma}_B+\frac{\hbar}{2}\mathds{1}_2)) + \ln(2\pi\hbar)$. The state now has one fewer mode, and the number of weights and means is now $(n_1+1)^2$. The process of post-selection is repeated until there is a single mode left with a total number of $n_w=\prod_{i=1}^{N-1}(n_i+1)^2$ weights and means, corresponding to a linear combination of $n_w$ Gaussians with index set $\mathcal{J}=\mathcal{J}_1\times\dots\times\mathcal{J}_{N-1}$.

%The quadratic scaling in the photon number is the same as simulating the density matrix in Fock basis up to cutoff $n$.

The heralding probability $p(\vb*{n})$ can be found by performing a summation over the final weights of the single-mode output state, 
\[
%p(\vb*{n}) = (2\pi\hbar)^{N-1}\sum_{k\in \mathcal{K}} e^{w_k}
p(\vb*{n}) = \sum_{j\in \mathcal{J}} e^{\tilde{w}_j}
\]
Calculating the probability can be troublesome because rounding errors can occur when using finite floating point precision. SciPy's \texttt{logsumexp} function \cite{virtanen_scipy_2020} utilizes shifted formulas to avoid overflow and reduce the chance of harmful underflow when computing $\log (p(\vb*{n}))$. However, the shifted \texttt{logsumexp} function is still sensitive to rounding errors due to floating-point arithmetic \cite{blanchard_accurate_2019}, and this affects the numerical stability of our simulations. As we shall see in the next section, several other relevant measures, such as the overlap and the characteristic function revolve around computing sums of exponential functions.

\subsection{State characterization}
\label{sec:state_character}
The quality of the output single-mode state from the GBS device can be evaluated in several ways. Its fidelity to another state, for example, a finite-energy GKP state with a Gaussian envelope, can be computed. Another way to determine the closeness of a non-Gaussian state to a GKP state is to compute the effective squeezing \cite{duivenvoorden_single-mode_2017}, which relates the GKP stabilizer expectation values to the variances of each Gaussian peak. A closely related GKP quality measure is the nonlinear GKP squeezing \cite{marek_ground_2024}, which relates the expectation value of an Hermitian operator consisting of linear combinations of grid displacements to the closeness of the state to a particular GKP lattice. In the following, we will show that all of these measures fit within the LCoG framework. See Table \ref{tab:state_char} for an overview of the figures of merit described here. 
\begin{table}[]
\caption{State characterisation of the \texttt{State} object.}
\begin{ruledtabular}
\centering
\begin{tabular}{lp{60mm}}
%\\ \hline\hline
Function & Description   \\
\colrule
%\hline
\texttt{overlap} & Compute the overlap of two \texttt{State} objects via Eq.\ \eqref{eq:overlap}. \\
\texttt{char\_fun} & Evaluate the characteristic function in point $\vb*{\alpha}$ via Eq.\ \eqref{eq:char_fun}. \\
\texttt{effective\_sqz} & Evaluate the effective squeezing in Eq.\ \eqref{eq:eff_sqz} via the characteristic function.\\
\texttt{gkp\_squeezing} & Evaluate the nonlinear GKP squeezing in Eq.\ \eqref{eq:gkp_squeezing} via the characteristic function.\\
%\hline\hline
\end{tabular}
\end{ruledtabular}
\label{tab:state_char}
\end{table}

\subsubsection*{Overlap}
The overlap between two states $\hat{\rho}_1$ and $\hat{\rho}_2$ can be computed via the phase-space integral over the product of the Wigner functions,
\[
\mathrm{Tr}[\hat{\rho}_{1}\hat{\rho}_{2}]  =(2\pi \hbar)^N \int d\vb*{q} W_{\hat{\rho}_{1}}(\vb*{q})W_{\hat{\rho}_{2}}(\vb*{q}),
\]
which for states that are LCoG, $W_{\hat{\rho}_{1}}(\vb*{q})=\sum_{k\in\mathcal{K}}e^{c_{k}}G_{\vb*{\mu}_{k},\vb*{\sigma}_{k}}(\vb*{q})$, and $W_{\hat{\rho}_{2}}(\vb*{q})=\sum_{l\in\mathcal{L}}e^{d_l}G_{\vb*{\nu}_{l},\vb*{\omega}_{l}}(\vb*{q})$, involves calculating $n\times m$ Gaussian overlaps where $n$ is the number of elements in $\mathcal{K}$ and $m$ is the number of elements in $\mathcal{L}$. This evaluates to a sum over $n\times m$ exponential functions,
\[
\mathrm{Tr}[\hat{\rho}_{1}\hat{\rho}_{2}]& =(2\pi \hbar)^N\sum_{kl}e^{c_{k}+d_{l}} \int  d\vb*{q} G_{\vb*{\mu}_{k},\vb*{\sigma}_{k}}(\vb*{q})G_{\vb*{\nu}_{l},\vb*{\omega}_{l}}(\vb*{q})\nonumber\\
&=
(2\pi \hbar)^N\sum_{kl}e^{c_{k}+d_{l}}G_{\vb*{\mu}_{k},\vb*{\sigma}_{k}+\vb*{\omega}_{l}}(\vb*{\nu}_{l})\nonumber\\
&=\sum_{kl} e^{c_k+d_l+\gamma_{kl}+\delta_{kl}}\label{eq:overlap}
\]
where $\gamma_{kl}=-\frac{1}{2}(\vb*{\mu}_{k}-\vb*{\nu}_{l})^T(\vb*{\sigma}_{k}+\vb*{\omega}_{l})^{-1}(\vb*{\mu}_{k}-\vb*{\nu}_{l})$ and $\delta_{kl}=N\ln(2\pi \hbar)-\frac{1}{2}\ln(2\pi \det(\vb*{\sigma}_{k}+\vb*{\omega}_{l}))$. If one of the states is pure, the overlap is equal to the quantum fidelity. The purity of a state can be obtained by computing the overlap with itself. 

\subsubsection*{Effective squeezing}
The effective squeezing \cite{duivenvoorden_single-mode_2017} is defined using the absolute value of the expectation value of the GKP stabilizer in the orthogonal quadrature. That is, for the effective squeezing in $x$, $\Delta_x$, one uses the $p$ stabilizer, which is a displacement in $p$, and vice versa,
\[
&\Delta_{q} = \sqrt{\frac{-2}{\abs{\alpha_q}^2}\ln(\lvert\langle\hat{D}(\alpha_q)\rangle\rvert)}\label{eq:eff_sqz},
\]
$\Delta_q$ can be related to a squeezing in decibel units with the following conversion $\Delta_{\text{dB}}=-10\log_{10}(\Delta ^2)$.

\subsubsection*{GKP nonlinear squeezing}
The GKP nonlinear squeezing \cite{marek_ground_2024} is defined as the expectation value of the operator $\hat{Q}$,
\[
\xi=\frac{1}{2}\langle\hat{Q}\rangle, \label{eq:gkp_squeezing}
\]
where $\hat{Q}$ is a linear combination of grid displacements. The operator is not invariant under Gaussian operations, and can therefore differentiate between logical states, which differ by a half-grid displacement in $x$. For example, for the logical GKP qubit in the computational basis with $j\in\{0,1\}$, the operator is
\[
\hat{Q}_{j}& =4\hat{\mathds{I}}-\hat{D}\left(\alpha\right)-\hat{D}\left(-\alpha\right)\nonumber\\
&-(-1)^{j}\hat{D}\left( i\frac{\alpha}{2} \right)
-(-1)^{j}\hat{D}\left( -i\frac{\alpha}{2}  \right),
\]
with $\alpha=\sqrt{2\pi}$. The nonlinear GKP squeezing operator for the qunaught state can be obtained by squeezing $\hat{Q}_j$, i.e. scaling the displacements by a factor of $\sqrt{2}^{-1}$ in $x$ and $\sqrt{2}$ in $p$, resulting in a symmetric grid with a one-dimensional code-space,
\[
\hat{Q}_{sj}&=4\hat{\mathds{I}}-\hat{D}\left(\frac{\alpha}{\sqrt{2}}\right)-\hat{D}\left(-\frac{\alpha}{\sqrt{2}}\right)\nonumber\\
&-(-1)^{j}\hat{D}\left( i\frac{\alpha}{\sqrt{2}} \right)-(-1)^{j}\hat{D}\left( -i\frac{\alpha}{\sqrt{2}} \right).
\]
The conversion to decibel units is $\xi_{\text{dB}}=-10\log_{10}(\xi)$. 

\subsubsection*{Characteristic function}
Both the effective squeezing and the GKP nonlinear squeezing involve evaluating the expectation values of displacement operators. This is the definition of the characteristic function, which for states in the form Eq.\ \eqref{eq:Wigner_sumofG} is a LCoG,
\[
\Tr[\hat{\rho}\hat{D}(\vb*{\alpha})]=\chi_{\hat{\rho}}(\vb*{\alpha})=\sum_{m\in\mathcal{M}}e^{w_{m}+i\vb*{\mu}_{m}^T\vb*{\Omega}\vb*{\alpha}-\frac{1}{2}\vb*{\alpha}^T\vb*{\Omega}\vb*{\sigma}_m\vb*{\Omega}^T\vb*{\alpha}}, \label{eq:char_fun}
\]
where $\vb*{\Omega}=\begin{pmatrix}
    0 & 1\\-1 &0
\end{pmatrix}$ is the symplectic form and $\vb*{\alpha}=(\Re \alpha,\Im \alpha)^T$ is the vectorized phase-space displacement. The effective squeezing in Eq.\ \eqref{eq:eff_sqz} and the nonlinear GKP squeezing in Eq.\ \eqref{eq:gkp_squeezing} are evaluated for a specific $\vb*{\alpha}$ coordinate, which involves calculating the sum of exponential functions in Eq.\ \eqref{eq:char_fun}.

\subsection{Analytical gradients with respect to parameters of the Gaussian circuit}\label{sec:gradients}

The Gaussian circuit, parameterized by the squeezing and beam splitter parameters in Fig.\ \ref{fig:GBS_circuits}, can be optimized to herald a specific target state for a given photon number pattern. We can choose the cost function for the optimization to be the quantum fidelity to a target state, or in the case of heralding a GKP state, we choose the effective squeezing or nonlinear GKP squeezing from Sec.\ \ref{sec:state_character}. In the LCoG framework, the cost function $h(g(\vb*{\theta}))$ will be typically be related to a sum over exponential functions,
% $\mathcal{N}=\sum_{m\in\mathcal{M}}e^{w_m}$ is the norm of the state, i.e. the measurement probability if the state is heralded. 
\[
g(\vb*{\theta})=\sum_{j\in\mathcal{J}}e^{f_j(\vb*{\theta})},
\]
where $\vb*{\theta}$ is a vector of Gaussian circuit parameters. The cost function could for example be the effective squeezing, which is a function of the characteristic function, $\Delta_q(\chi(\alpha_q))$. In the following, $g(\vb*{\theta})$ may be the overlap in Eq.\ \eqref{eq:overlap} or the characteristic function in Eq.\ \eqref{eq:char_fun}. It can be beneficial to provide analytical gradients of the cost function to the optimizer. The gradient of $g$ with respect to a parameter $\phi=\theta_i$ of the Gaussian circuit can be computed via the chain rule for partial derivatives, 
\[
\frac{\partial g(\vb*{\theta})}{\partial \phi}=\frac{\partial \sum_{j}e^{f_{j}(\vb*{\theta})}}{\partial \phi}=\sum_{j}\frac{\partial f_{j}(\vb*{\theta})}{\partial \phi}e^{f_{j}(\vb*{\theta})} .\label{eq:gradient_chain_rule}
\]
In the following, we will discuss how to calculate the partial derivative of the exponential function.

\subsubsection*{Gradients of the characteristic function}
The characteristic function appears in both the effective squeezing and the nonlinear GKP squeezing. For the characteristic function, the exponent in Eq.\ \eqref{eq:gradient_chain_rule} is
\[
f_{j}(\vb*{\theta})=w_{j}+\gamma_{j}(\vb*{\theta})+i\vb*{\tilde{\mu}}_{j}^T(\vb*{\theta})\vb*{\Omega}\vb*{\alpha}-\frac{1}{2}\vb*{\alpha}^T\vb*{\Omega}\vb*{\tilde{\sigma}}(\vb*{\theta})\vb*{\Omega}^T\vb*{\alpha}.
\]
Here, we explicitly separate the log-weights, $w_j$, which arise from the PNRD approximation in Eq.\ \eqref{eq:weights_fock_povm}, and the partial Gaussian overlaps, $\gamma_j$, computed via Eq.\ \eqref{eq:gamma_sum_of_G} due to the partial measurements, the latter of which have a $\vb*{\theta}$ dependency. The partial derivative of each exponent is therefore,
\[
\frac{\partial f_{j}}{\partial \phi}=\frac{\partial \gamma_{j}}{\partial \phi}+\frac{i\partial \vb*{\tilde{\mu}}_{j}^T}{\partial \phi}\vb*{\Omega}\vb*{\alpha}-\frac{1}{2}\vb*{\alpha}^T\vb*{\Omega}\frac{\partial \vb*{\tilde{\sigma}}}{\partial \phi}\vb*{\Omega}^T\vb*{\alpha}.
\] 
The partial derivatives of $\gamma_j$, $\tilde{\vb*{\mu}}_j$, and $\vb*{\tilde{\sigma}}$ can be evaluated using only the partial derivatives of the initial multimode Gaussian state in Eq.\ \eqref{eq:init_gauss_state}. The explicit forms of these partial derivatives are given in Eq.\ \eqref{eq:partial_weight}, \eqref{eq:partial_mean}, and \eqref{eq:partial_cov} in App. \ref{app:partial}. They solely involve elements of the initial gradient of the covariance matrix, 
\[
\frac{\partial\vb*{\sigma}_1}{\partial \phi} =\frac{\hbar}{2}\vb*{X}\left(  \frac{\partial\vb*{S}}{\partial \phi}\vb*{S}^T + \vb*{S} \frac{\partial\vb*{S}^T}{\partial \phi}\right)\vb*{X}^T, \quad \frac{\partial \vb*{\mu}_1}{\partial\phi}=\vb*{0}_{2N}.
\]
As each mode is projected onto an approximate Fock state, the gradients of the overlap factors, displacement vectors and covariance matrix can be calculated similarly to how the covariance matrix, displacement vector and overlap factor are calculated as a result of a partial measurement in Sec.\ \ref{sec:sumofG}. The gradient of the covariance matrix and displacement vector is partitioned into the mode being measured (mode labeled by $B$) and the rest of the system (modes labeled by $A$),
\[
\frac{\partial \vb*{\sigma}_1}{\partial \phi} = \begin{pmatrix}
\frac{\partial\vb*{\sigma}_{A}}{\partial \phi} & \frac{\partial\vb*{\sigma}_{AB}}{\partial \phi} \\
\frac{\partial\vb*{\sigma}_{BA}}{\partial\phi} & \frac{\partial\vb*{\sigma}_{B}}{\partial\phi}
\end{pmatrix}\qand
\quad \frac{\partial\vb*{\mu}_1}{\partial \phi}=\begin{pmatrix}
\frac{\partial\vb*{\mu}_{A}}{\partial \phi}  \\
\frac{\partial\vb*{\mu}_{B}}{\partial \phi}
\end{pmatrix}.
\]
After the measurement of mode $B$, the gradients in modes $A$, $\frac{\partial\vb*{\sigma}_{A}'}{\partial \phi}$, $\frac{\partial\vb*{\mu}_{j,A}'}{\partial \phi}$, and $\frac{\partial\gamma_j}{\partial \phi}$ can be found using the transformation rules in App. \ref{app:partial}. Three additional attributes related to the gradients are therefore added to the \texttt{State} object, as listed in Table \ref{tab:state_partial}, if the gradients are requested. When gradients are requested, the simulation complexity scales linearly with the number of circuit parameters. 

\begin{table}[]
\caption{Attributes related to gradients of the \texttt{State} object. $n_g$ is the number of gradients, i.e. number of circuit parameters, $n_w$ is the number of weights, and $N$ is the number of modes.}
\centering
\begin{ruledtabular}
\begin{tabular}{lp{30mm}p{25mm}}
%\\ \hline\hline
Attribute & Description & Dimension \\
\colrule
%\hline
\texttt{partial\_log\_weights} & List of log-weight gradients & $(n_g,n_w,1)$ \\
\texttt{partial\_means} & List of displacement vector gradients & $(n_g,n_w,2 N)$ \\
\texttt{partial\_covs} & List of covariance matrix gradients & $(n_g,n_w,2 N,2N)$ or $(n_g,1,2 N,2N)$ \\
%\hline\hline
\end{tabular}
\end{ruledtabular}
\label{tab:state_partial}
\end{table}

\subsubsection*{Gradients of the fidelity}
When evaluating the gradients of the quantum fidelity in Eq.\ \eqref{eq:overlap} with respect to a pure state in the form of Eq.\ \eqref{eq:Wigner_sumofG}, each exponent in the overlap function in Eq.\ \eqref{eq:gradient_chain_rule} is
\[
f_{kl}(\vb*{\theta})=w_k+\gamma_k(\vb*{\theta})+d_l +\gamma_{kl}(\vb*{\theta})+\delta_{kl}(\vb*{\theta}),
\]
where the $k$ index is for the heralded state and the $l$ index is for the target state with log-weights $d_l$. Once again, we separate the weights $w_k$ of the PNRD and the Gaussian overlaps $\gamma_k$ due to the partial measurements. $\gamma_{kl}$ is the Gaussian overlap factor in the fidelity calculation. The partial derivative of $f_{kl}$ is given by
\[
\frac{\partial f_{kl}}{\partial \phi}=\frac{\partial\gamma_{k}}{\partial \phi}+\frac{\partial\gamma_{kl}}{\partial \phi}+\frac{\partial\delta_{kl}}{\partial \phi}
\]
The partial derivatives of $\gamma_k$, $\gamma_{kl}$ and $\delta_{kl}$ are given in Eq.\ \eqref{eq:partial_weight}, \eqref{eq:partial_overlap_gamma}, and \eqref{eq:partial_overlap_delta}. 

\subsubsection*{Gradients of the success probability}
Note that the heralded state is normalized by the success probability, which is also dependent on the circuit configuration, and therefore contributes to the partial derivative of the cost function. We make the following substitution in the exponential argument to include the norm $f_{j}(\vb*{\theta})\mapsto f_{j}(\vb*{\theta})-\ln(p(\vb*{n},\vb*{\theta}))$. The partial derivative becomes,
\[
\frac{\partial f_{j}}{\partial \phi}\mapsto \frac{\partial f_{j}}{\partial \phi} - \frac{1}{p(\vb*{n})} \frac{\partial p(\vb*{n})}{\partial \phi},
\]
where the partial derivative of the norm
\[
\frac{\partial p(\vb*{n})}{\partial \phi} = \sum_{j} \frac{\partial \gamma_{j}}{\partial \phi} e^{w_{j}}
\]
which only depends on the partial derivative of the weights, i.e. the Gaussian overlap factors $\gamma_j$ due to the measurements.

\subsection{Mapping to the minimal LCoGs}
\label{sec:state_reduce}
When simulating photon-number measurements on several modes, the exponential growth in the number of Gaussians becomes a limiting factor. In the heralded output state described in Sec.\ \ref{sec:output_state}, we use $\prod_i(n_i+1)^2$ Gaussians, where $\vb*{n}=(n_1,\dots,n_{N-1})^T$ is the photon-number pattern measured in the rest of the modes. Eventually, the system runs out of memory for storing millions of Gaussians. In this section, we propose a method, called  \texttt{rank\_reduce}, which reduces the number of Gaussians to the minimal number required to capture the non-Gaussian features.

The idea behind the reduction method is related to the stellar rank of CV quantum states \cite{chabaud_stellar_2020}. The single-mode output state $\ket{\psi_{\vb*{n}}}$ of a lossless $N$-mode GBS device is a displaced, squeezed superposition of at most $r=\sum_i n_i$ photons \cite{su_conversion_2019}, 
\[
\ket{\psi_{\vb*{n}}}=D(\alpha)\hat{S}(z)\underbrace{\sum_{n=0}^{r} c_n \ket{n}}_{\ket{\psi_\text{core}}}.\label{eq:GBS_out_fock_pure}
\]
$r$ is the stellar rank of the single-mode state, defined by the bounded support over the Fock basis of the core state. A LCoG representation of the state $\ket{\psi_{\vb*{n}}}$ that uses only $\left(r +1\right)^2$ Gaussians must exist, since the core state can be expressed as a superposition of $r+1$ coherent states as described in Sec.\ \ref{sec:coherent_state_decomp}. 

In the following, the \texttt{rank\_reduce} method will be described first for the lossless case, and then generalized to the situation where the output state is mixed. If at any point in the simulation there is a single-mode state, the \texttt{rank\_reduce} method can be used to re-express the state more compactly, allowing projection onto higher photon numbers and the simulation of additional modes in a cascaded interferometer topology.

\subsubsection*{Mapping of a pure state}
\begin{figure}
    \centering
    \includegraphics[width=\linewidth]{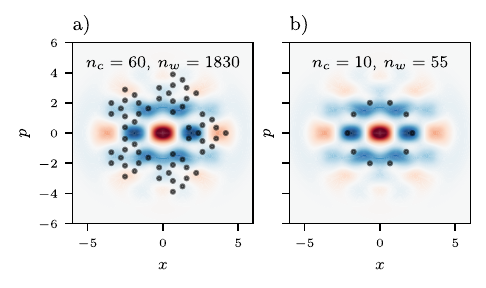}
    \caption{\textbf{Reduction by stellar rank}. The Wigner function of the output state of a random four mode GBS circuit (Clements topology) heralded by the photon pattern $\vb*{n}=(4,3,2)^T$ before a) and after b) the \texttt{rank\_reduce} method is applied. The black points are the centers of the coherent states in the superposition. In a) the centers of the points are scaled by a factor of 4 for better visualization. The number of coherent states in the superposition is $n_c$ and the number of Gaussians in the red-LCoG formalism used here is $n_w$. }
    \label{fig:rank_reduce}
\end{figure}
The Wigner function of the single-mode output state from the GBS is a linear combination of $n_w=\prod_i(n_i+1)^2$ Gaussians,
\[
W_{\hat{\rho}}(\vb*{q})&=\sum_{j\in\mathcal{J}} e^{w_j} G_{\vb*{\mu}_{j},\vb*{\sigma}}(\vb*{q})\label{eq:GBSout1},
% &=\sum_{k=1}^{k_1}{w_{k}}G_{\vb*{\mu}_{k},\vb*{\sigma}}(\vb*{q}) 
% + 2\sum_{k=k_1+1}^{k_2}\Re\left[w_k G_{\vb*{\mu}_{k},\vb*{\sigma}}(\vb*{q})\right].\label{eq:GBSout2}
\]
where $\mathcal{J}$ is a set of indices.  Since the covariance matrix is the same for each term, we can perform the Williamson decomposition, $\vb*{\sigma}= \vb*{S}\vb*{D}\vb*{S}^T=\vb*{S}(\frac{\hbar}{2}\mathds{1}_{2}+\vb*{\nu})\vb*{S}^T$, resulting in a symplectic matrix $\vb*{S}$ and additional thermal eigenvalue contributions $\vb*{\nu}$ if the state is mixed. In the pure state case, $\vb*{\nu} = 0_{2,2}$ is a zero matrix.

The Gaussian operation with symplectic matrix $\vb*{S}$ can be removed by performing the inverse symplectic operation, resulting in a new state $\hat{\sigma}=\hat{U}_{\vb*{S}}^\dagger \hat{\rho}\hat{U}_{\vb*{S}}$, where $\hat{U}_{\vb*{S}}$ is a Gaussian unitary operation with symplectic matrix $\vb*{S}$,
\[
W_{\hat{\sigma}}(\vb*{q})&=\sum_{j\in\mathcal{J}} e^{w_j} G_{\vb*{\mu}_{j}', \frac{\hbar }{2}\mathds{1}_{2}}(\vb*{q})\label{eq:Wig_noS}.
\]
The transformed means have been re-labeled with primes,  $\vb*{S}^{-1}\vb*{\mu}_{j}\mapsto\vb*{\mu}_{j}'$, and the weights remain unchanged after the Gaussian operation. Note that the density operator $\hat{\sigma}$ can now be written as a linear combination of coherent state outer products, 
\[
\hat{\sigma}&=\sum_{j\in\mathcal{J}} e^{w_j-d_j}\ketbra{\alpha_{j}}{\beta_j} \label{eq:sigma_full},
% &=\sum_{j=1}^{k_1} w_j \ketbra{\alpha_j}{\alpha_j} + \sum_{j={k_1+1}}^{k_2} \left(\frac{w_j}{d_j}\ketbra{\alpha_j}{\beta_j} + \frac{w_j^*}{d_j^*}\ketbra{\beta_j}{\alpha_j} \right)
\]
where the coherent state amplitudes $\alpha_j$ and $\beta_j$ as well as $d_j$ can be obtained from the $\vb*{\mu}_j$'s via Eq.\ \eqref{eq:mu_to_alpha_coherent}. In the next step, we can convert the state into the Fock basis via Eq.\ \eqref{eq:coherent_to_fock_dm},
\[
\hat{\sigma}= \sum_{mn}\rho_{mn}\ketbra{m}{n} \label{eq:sigma_fock_dm}.
\]
We would like to find all the photon-number state vector coefficients, $c_n$ in Eq.\ \eqref{eq:GBS_out_fock_pure} from the Fock tensor $\rho_{mn}=c_m c_n^*$. By considering the following ratio
$\rho_{nr}/\rho_{rr}=c_n c_r^* /\abs{c_r}^2=c_n/c_r$, we realize that only the $\rho_{nr}$ elements where $n\in\{0,\dots,r\}$ need to be computed \cite{su_conversion_2019}. After obtaining the photon-number state vector, we can map the core state to $r+1$ coherent states by using the coherent state decomposition of a superposition of photon-number states in Eq.\ \eqref{eq:fock_superpos_coherent}. The Wigner function of this superposition of coherent states has $(r+1)^2$ Gaussians, which can be computed using the methods from Sec.\ \ref{sec:sumofG} and \ref{sec:coherent_state_decomp}. An example of the Wigner function of a heralded non-Gaussian state before and after applying the \texttt{rank\_reduce} method is plotted in Fig.\ \ref{fig:rank_reduce}. The Wigner functions look identical, but the state in b) is represented with significantly fewer Gaussians.
% This reduction algorithm is implemented in the \texttt{reduce.py} module using broadcasting where appropriate to speedup the summation over Gaussian weights. 

\subsubsection*{Mapping for mixed states}
For mixed states, the thermal contributions from the Williamson decomposition are non-zero, $\vb*{\nu}\neq 0_{2,2}$. After removing the Gaussian operation with symplectic matrix $\vb*{S}$,
\[
% W_{\hat{U}^\dagger\hat{\rho}\hat{U}}(\vb*{q})&=\sum_{k}w_{k}G_{\vb*{S}^{-1}\vb*{\mu}_{k}, \frac{\hbar }{2}\mathds{1}_{2}+\vb*{\nu}}(\vb*{q})+2 \sum_{l} \Re\left[ w_{l}G_{\vb*{S}^{-1}\vb*{\mu}_{l}, \frac{\hbar }{2}\mathds{1}_{2}+\vb*{\nu}}(\vb*{q})\right]\\
W_{\hat{\sigma}}(\vb*{q})&=\sum_{j\in\mathcal{J}}e^{w_j}G_{\vb*{\mu}_{j}', \frac{\hbar }{2}\mathds{1}_{2}+\vb*{\nu}}(\vb*{q}) \label{eq:Wig_noS}
\]
The additive $\vb*{\nu}$ matrix can be interpreted as having originated from a random Gaussian displacement channel \cite{quesada_quadratic_2022},
\[
W_{\hat{\sigma}}(\vb*{q})&=\int  \dd{\vb*{\alpha}}^2 P(\vb*{\alpha})W_{\hat{D}(\alpha)\hat{\sigma}'\hat{D}^{\dagger}(\alpha)}(\vb*{q}),
\]
where $P(\vb*{\alpha})=G_{\vb*{0},\vb*{\nu}}(\vb*{\mu}_{\alpha})$ is a Gaussian distribution centered around zero with variance $\vb*{\nu}$, and $\vb*{\mu}_\alpha =\sqrt{2\hbar}(\Re(\alpha),\Im(\alpha))^T$. We now proceed with $\hat{\sigma}'$, which is an unphysical state where the thermal eigenvalue matrix $\vb*{\nu}$ has simply been subtracted from the covariance matrix of $\hat{\sigma}$, 
\[
W_{\hat{\sigma}'}(\vb*{q})=\sum_{j\in\mathcal{J}}e^{w_j}G_{\vb*{\mu}'_{j},\frac{\hbar}{2}\mathds{1}_{2}}(\vb*{q})
\]
This state is not a pure state, because the log-weights still contain elements of the initial lossy covariance matrix. However, $\hat{\sigma}'$ can still be written as a linear combination of coherent state outer products using Eq.\ \eqref{eq:sigma_full}. We can write this unphysical state in the Fock basis via Eq.\ \eqref{eq:sigma_fock_dm}, truncate the expansion to some value $r'$, and convert the state from the Fock basis back to $(r'+1)^2$ coherent states. This process is described in more detail in App. \ref{app:coherent_to_fock}. Lastly, the thermal contributions $\vb*{\nu}$ can then be added back to the vacuum covariance matrix, and the symplectic operation can be re-applied, arriving at the original state, but now with fewer Gaussians. 

The photon-number cutoff $r'$ can be set via Chebyshev's inequality. Given the mean $\mu_n = \langle\hat{n}\rangle$ and variance $\sigma_n = \langle\hat{n}^2\rangle -\mu_n^2$ of the photon number of $\hat{\sigma}'$, which for LCoGs can be found via Eq.\ \eqref{eq:photon-number_mean_sumG} and  Eq.\ \eqref{eq:photon-number_variance_sumG}, we can bound the probability of the photon number deviating from the mean by more than $k\sqrt{\sigma_n}$ by $1/k^2$,
% the probability of photon numbers  $\mu_n + k\sqrt{\sigma_n}$ is bounded by $1/k^2$,
\[
P(|n-\mu_n| \geq k\sqrt{\sigma_n})\leq\frac{1}{k^2}.
\]
We set the photon number cutoff to be $r'=\lceil\mu_n+k\sqrt{\sigma_n}\rceil$ where the number of standard deviations $k$ is a free parameter. The reduction procedure for mixed states is more computationally demanding than the reduction of pure states, because a sum over the (truncated) density matrix must be computed for each weight. By using symmetries, such as $\rho_{mn}=\rho_{nm}^*$, and broadcasting, we can speed up the summations slightly. The removal of $\vb*{S}$ and $\vb*{\nu}$ also ensures that the extension in the photon number basis is kept to a minimum. The truncation of the photon number re-introduces the possibility of propagating errors if it is insufficient, which can be countered by setting $k$ to be conservatively large. This results in a less significant reduction in the number of Gaussians, but it will always provide a speedup, unless we pick $k$ so large that $r'\sim \prod_{i} n_i$. 

In this section, the \texttt{rank\_reduce} method has been described for single-mode states. The stellar rank formalism can be extended to multi-mode states \cite{chabaud_classical_2021}, in which the multi-mode core state is also bounded in the Fock basis. It should therefore be possible to generalize the reduction procedure to multi-mode states that are both pure and mixed.

% Following \cite{su_conversion_2019}, finding the relative Fock coefficients $\frac{c_{n_{r}}^*c_{m}}{\lvert  c_{n_{r}}\rvert^2 }=\frac{c_{m}}{c_{n_{r}}}$ for all $m$ up to $n_r$ is sufficient to be able to describe the pure state in Fock basis up to a constant factor $c_{n_r}$, which will be handled by the normalisation. 

\subsection{Numerical stability and errors}
\label{sec:numerical_stability}
In this section, we discuss the sources and propagation of errors when simulating the single-mode output state of the Gaussian Boson sampling device in Fig \ref{fig:GBS_circuits} in the LCoG framework. By partially projecting onto $m$ approximate Fock states with error $\epsilon$, the fidelity of the output state will be $(1-\epsilon)^{m}$. Marshall and Anand \cite{marshall_simulation_2023} argue that in order to achieve a total fidelity $1-\delta$, one can simply scale the initial error linearly, $\epsilon\approx \delta /m$. Similarly, when performing the \texttt{rank\_reduce} operation (on a pure state), there is an additional error contribution $\epsilon_r$ from the chosen final radius of the coherent state ring, and the total fidelity of the approximation becomes $(1-\epsilon)^m(1-\epsilon_r)$. In principle, the error can be made arbitrarily small. However, in the numerical implementation, small infidelities cause rounding errors when using finite floating point precision. This was shown in Fig.\ \ref{fig:fock_approx}, where for small radii, the fidelity of the Fock approximation with coherent states moves away from the theoretical line. The source of this error is the inability to accurately compute the norm of the state, i.e. the summation over the weights. In both the coherent state and thermal state decomposition, the norm is inversely related to the fidelity of the Fock approximation. When precision of the norm becomes close to the smallest float, i.e. $\sim2.22\times 10^{-16}$ for \texttt{float64} in Python, problems arise because the norm cannot be accurately computed. Telltale signs of the onset of numerical instabilities include negative probabilities, or overlaps that are greater than 1. Depending on the application, we consider infidelity $\epsilon$ settings between $10^{-8}$ and  $10^{-4}$ to be a safe choice for most photon numbers. 

% Propagation of error with several fock measurements  (multiplicative). 
% There are no problems associated with defining an insufficient cutoff. The error of the approximation due to the finite ring in phase space can, in principle, be made arbitrarily small. However, the implementation of the framework with finite floating point precision can cause numerical instabilities that can be difficult to identify. Some tell-tale signs of problems are negative probabilities and overlaps and purities greater than 1. 

\section{Application: Optimal GBS circuits for qunaught state preparation}
\label{sec:results}
To demonstrate the utility of the LCoG simulation framework with the new PNRD capabilities described in  Sec.\ \ref{sec:lcog_simulation}, we built an optimizer for Gaussian Boson sampling circuits as shown in Fig.\ \ref{fig:GBS_circuits}, with the target of heralding qunaught states \cite{duivenvoorden_single-mode_2017}, which are symmetric GKP states with an equal spacing of $\sqrt{2\pi\hbar}$ between peaks in both $x$ and $p$. The preparation of a fault-tolerant qunaught state is a critical bottleneck in the development of a photonic quantum computer. These states are more difficult to prepare with GBS because they require at least four peaks to be fault-tolerant, compared to only three peaks for the logical state with $2\sqrt{\pi\hbar}$ spacing in $x$ (see S.M. of \cite{aghaee_rad_scaling_2025} for a more detailed discussion), which requires a larger GBS circuit. 

Additionally, the optical components and photon detectors can have a quantum efficiency below one, which will affect the quality of the heralded state. In this section, we show that the LCoG methodology is well-suited for the optimization of state preparation circuits affected by optical losses.

\subsection{Optimization settings}
For the results shown in this section, the cost function was chosen to be the sum of the effective squeezing in $x$ and $p$: $h(\vb*{\theta}) = \Delta_x(\vb*{\theta})+\Delta_p(\vb*{\theta}).$
The gradient of the effective squeezing with respect to a Gaussian parameter $\phi=\theta_i$ can be found via the chain rule,
\[
\frac{\partial\Delta_{q}}{\partial \phi} = - \frac{1}{2}\Delta_q^{-\frac{1}{2}}\frac{1}{\lvert \alpha_{q} \rvert^2 } \frac{\chi(\alpha_{q})}{\lvert \chi(\alpha_{q}) \rvert^2} \frac{\partial\chi(\alpha_{q})}{\partial \phi}
\]
where the gradient of the characteristic function $\partial\chi(\alpha_q)/\partial\phi$ is given in Section \ref{sec:gradients}. Other figures of merit discussed in Section \ref{sec:state_character}, such as the fidelity, are also compatible.
% \[
% \frac{\partial\chi(\alpha_{q})}{\partial \phi} &= \frac{1}{\mathcal{N}}\sum_{m\in\mathcal{M}}\frac{\partial f_m}{\partial\phi} e^{f_m} - \frac{1}{\mathcal{N}}\chi(\alpha_{q})\frac{\partial\mathcal{N}}{\partial\phi}
% \]
% The heralding probability is also dependent on the circuit parameters, and therefore its gradient must also be included.
We optimize three different GBS circuit topologies shown in Fig.\ \ref{fig:GBS_circuits} without loss. From the literature \cite{takase_gottesman-kitaev-preskill_2023}, we know that GKP states can be synthesized using a process called coherent bifurcation, which can be implemented by interfering $x$ and $p$ squeezed states on phase-less beam splitters in the inverse-cascade configuration and projecting onto the same photon number $n$ in all detectors. We therefore set all the beam splitter and squeezing phases to zero to reduce the parameter space. This gives us a total of  $N(N+1)/2$ parameters for the rectangular/Clements topology, and $2N-1$ parameters for the cascades, where $N$ is the number of modes. Finding a global minimum with this many parameters is a nontrivial task, and we suspect that several different circuit configuration can produce a good grid state. Inspired by Tzitrin et al. \cite{tzitrin_progress_2020}, we employ the \texttt{basinhopping} algorithm from \texttt{scipy.optimize}, which stochastically probes the parameter space by iterating between a local minimization and a hop to a different area of the space. We find that the \texttt{L-BFGS-B} and \texttt{SLSQP} local minimization methods have the best performance, with the former being the fastest. We set the maximum squeezing to $r_{\text{max}}= 15$ dB \cite{vahlbruch_detection_2016}, i.e. we bound the squeezing parameters to be $r\in[-1.73,1.73]$. Since the numerical stability of our cost function is at risk when the heralding probability is too low, we bound the beam splitter parameters to $\theta\in\left[\frac{1}{10},\frac{\pi}{2}-\frac{1}{10}\right]$ to avoid configurations where e.g. none of the light goes into the detectors. We pick the starting guess from a uniform distribution. In general, the performance of the optimization depends on the initial starting point, so we perform $10$ optimization runs for each circuit type and photon-number pattern $n$ with different starting guesses and we set the number of iterations in \texttt{basinhopping} to $20$. In \texttt{scipy.optimize}, one can choose to provide analytical gradients when evaluating the cost function with the \texttt{"jac"=True} setting. The cost of evaluating the analytical gradients scales linearly with the number of parameters, $n_g$, so this can be an expensive cost function compared to the gradient-free method. If not provided, the gradients are approximated using finite differences, which requires many function evaluations. Each optimization is run twice; once providing the analytical gradients and once without.  

\begin{figure*}
    \centering
    \includegraphics[width=0.85\linewidth]
    {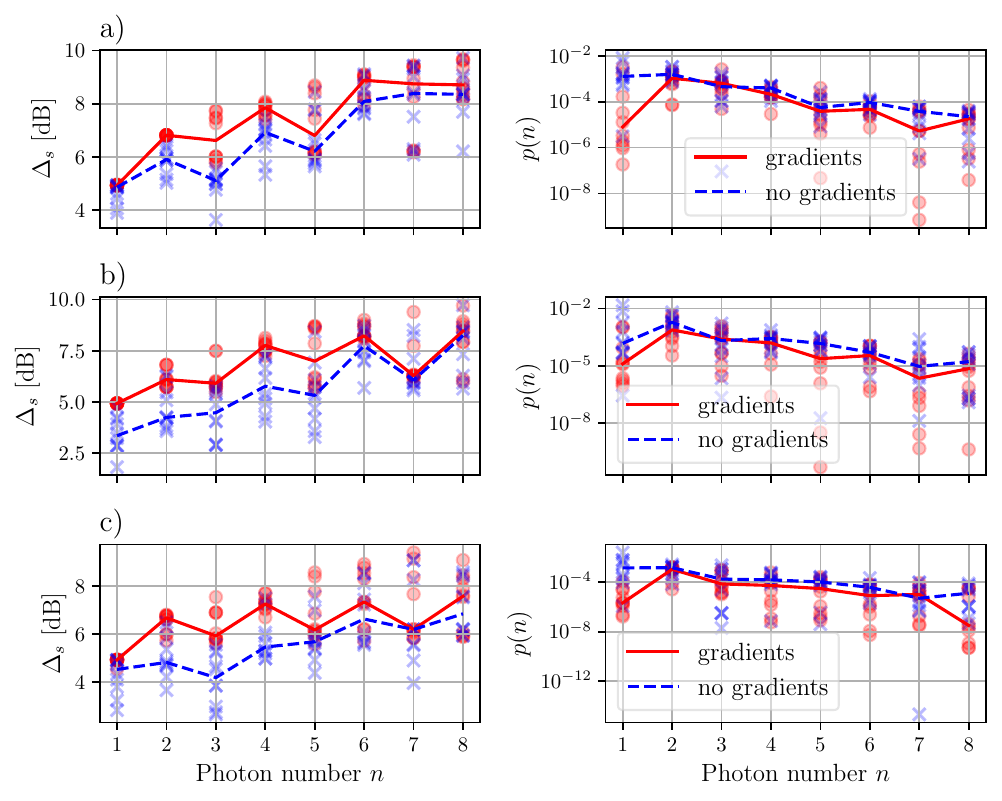}
    \caption{\textbf{Optimization of a four mode GBS circuit}. The symmetric effective squeezing $\Delta_s^2=\frac{1}{2}(\Delta_x^2+\Delta_p^2)$ \cite{aghaee_rad_scaling_2025} (left) and success probability $p(n)$ (right) of $10$ optimized solutions of the $N=4$ mode GBS circuits with  rectangular a), inverse cascade b), and cascade c) beam splitter arrangement with (red) and without (blue) the use of analytical gradients. The lines are the median $\Delta_s$ and $p(n)$.}
    \label{fig:opt_4_modes}
\end{figure*}
\begin{figure*}
    \centering
    \includegraphics[width=0.85\linewidth]
    {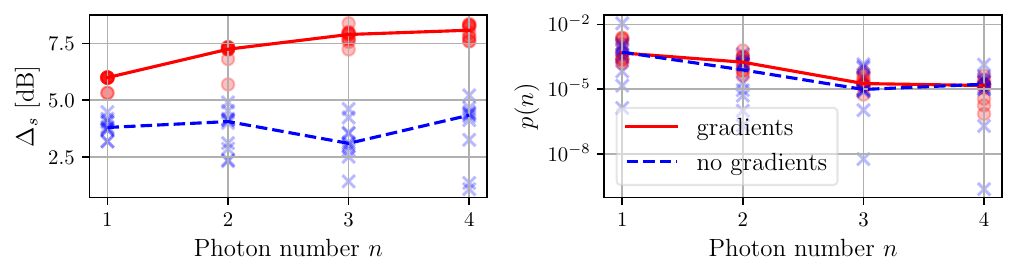}
    \caption{\textbf{Optimization of a five mode GBS circuit}. The symmetric effective squeezing $\Delta_s$ and success probability $p(n)$ of $10$ optimized solutions of the $N=5$ mode GBS circuits with a rectangular beam splitter topology with (red) and without (blue) the use of analytical gradients The lines are the median of 10 runs.}
    \label{fig:opt_5_modes}
\end{figure*}

\subsection{Results}

We show the results of the optimization runs for four and five modes in Fig.\ \ref{fig:opt_4_modes} and \ref{fig:opt_5_modes}. In general, the optimization benefits from using analytical gradients, especially when there are many parameters, as in the case of the five mode Clements circuit in Fig.\ \ref{fig:opt_5_modes}, which has 15 parameters. The gradient-free optimization has a tendency to get stuck on (two-peak) squeezed cat states, i.e. states with a large difference between $\Delta_x$ and $\Delta_p$. However, for higher photon-numbers, the difference in performance becomes smaller, likely because it is easier to find solutions which minimize $\Delta_x$ and $\Delta_p$ simultaneously. The success probability of the solutions found using analytical gradients is generally lower, but this is because the cost function only prioritizes the effective squeezing. The probability can be included in the cost function, for example with   $h=\frac{1}{2}(\Delta_x+\Delta_p) - c p(\vb*{n})$ where $c$ can be chosen freely \cite{tzitrin_progress_2020}. 

Circuits with the Clements beam splitter topology produce the best states, as they cover the full set of possible Gaussian operations (within the given parameter bounds). The inverse-cascade topology is able to generate the same type of solution, however the Clements decomposition "hits" them more often, producing better solutions on average. 

\begin{figure}
    \centering
    \includegraphics[width=\linewidth]{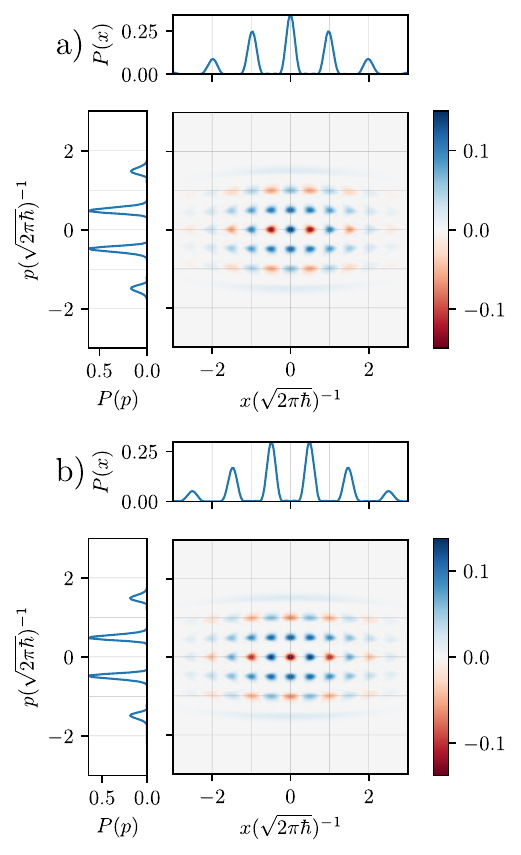}
    \caption{\textbf{Optimal qunaught states.} The Wigner function and marginal distribution of the qunaught states prepared by the optimal four-mode GBS circuit with parameters listed in Table \ref{tab:qunaught_opt_parameters} heralded by a  photon pattern of $n=(8,8,8)^T$ a) and $n=(9,9,9)^T$ b).}
    \label{fig:Wigner_opt}
\end{figure}

We highlight the two optimization results that are closest at producing a fault-tolerant qunaught state, a state with a symmetric effective squeezing of $\Delta_s\geq 9.75$ dB \cite{aghaee_rad_scaling_2025}, in Table \ref{tab:qunaught_opt_results} and Fig.\ \ref{fig:Wigner_opt}. The optimal parameters of the circuit are tabulated in Table \ref{tab:qunaught_opt_parameters}. In comparison, the synthesis of qunaught states with three ideal coherent bifurcations \cite{takase_gottesman-kitaev-preskill_2023}, corresponding to a four-mode GBS circuit, produces a state with $\Delta_s=8.62$ dB for $n=8$ and $\Delta_s=8.89$ dB for $n=10$, albeit at a higher success probability of $4.22\times10^{-5}$ and $2.19\times10^{-5}$, respectively. The numbers given here are from our own simulations of the coherent bifurcation process using the methodology developed in this work. In particular, we utilize the \texttt{rank\_reduce} method from Sec.\ \ref{sec:state_reduce} to reduce the number of Gaussians after each coherent bifurcation. It might be possible to model the optimized Clements circuit in Table \ref{tab:qunaught_opt_parameters} by three approximate coherent bifurcations followed by an inline squeezing operation to align the grid. 

\begin{figure*}
    \centering
    \includegraphics[width=0.85\linewidth]
    {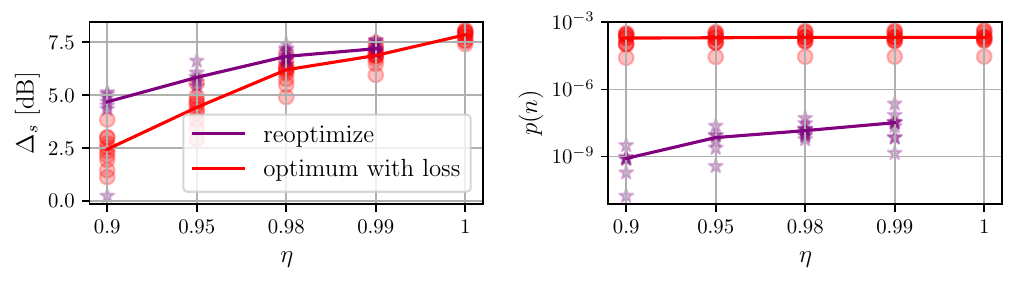}
    \caption{\textbf{Re-optimizing with loss}. The symmetric effective squeezing $\Delta_s$ and success probability $p(n)$ of the optimal solutions of the Clements configuration with photon pattern $(4,4,4)$ under a uniform loss channel with transmissivity $\eta$ (red) and the re-optimized solutions (purple) when the loss is included. Solid lines are the medians. }
    \label{fig:re-opt}
\end{figure*}

\begin{table}[]
\centering
\caption{Best results of the four mode GBS circuit in Fig.\ \ref{fig:GBS_circuits}b optimized to prepare a qunaught state heralded by photon pattern $(n,n,n)$.}
\label{tab:qunaught_opt_results}
\begin{ruledtabular}
\begin{tabular}{ccccc}
%\hline \hline
$n$ & Probability & $\Delta_x$ {[}dB{]} & $\Delta_p$ {[}dB{]} & $\Delta_s$ {[}dB{]}  \\ 
%\hline
\colrule
$8$ & $3.47\times10^{-5}$ & $8.35$ & $11.73$ & $9.72$ \\
$9$ & $7.67\times10^{-6}$ & $8.37$ & $12.38$ & $9.93$ \\
%\hline \hline
\end{tabular}
\end{ruledtabular}
\end{table}

\begin{table*}[]
\centering
\caption{Optimal circuit parameters for the preparation of the qunaught states shown in Fig.\ \ref{fig:Wigner_opt}. }
\label{tab:qunaught_opt_parameters}
\begin{ruledtabular}
\begin{tabular}{ccccccccccc}
%\hline \hline
$n$ & $r_1$ {[}dB{]} & $r_2$ {[}dB{]} & $r_3$ {[}dB{]} & $r_4$ {[}dB{]} & $\theta_1$ & $\theta_2$ & $\theta_3$ & $\theta_4$ & $\theta_5$ & $\theta_6$ \\ 
%\hline
\colrule
8 & -10.02 & -13.15 & -15.00 & 12.04 & 1.45 & 0.46 & 1.37 & 0.68 & 0.10 & 1.27 \\
9 & -8.20 & -11.52 & 12.22 & -12.96 & 1.02 & 0.95 & 0.74 & 0.74 & 0.23 & 1.46\\
%\hline \hline
\end{tabular}
\end{ruledtabular}
\end{table*}

\subsection{Optimizing circuits with loss }

Inefficient optical components can be modeled with the addition of a photon loss channel after the component. Similarly, the quantum efficiency of detectors can be modeled by applying a photon loss channel in front of the detector. The complexity of simulating quantum state preparation with the addition of these loss channels remains unchanged, because our methodology tracks the Wigner function, i.e. the density operator, and its complexity is dependent on the photon number pattern. As a proof of concept, we demonstrate that our optimizer can help mitigate the effects of optical losses. 

 We consider a 4-mode circuit in Fig.\ \ref{fig:GBS_circuits} with a rectangular beam splitter configuration, optimized to herald a qunaught state with the photon pattern $(4,4,4)$. We model PNRDs with quantum efficiency $\eta$, and re-optimize the circuit with a local optimization, using \texttt{scipy.minimize} and providing analytical gradients. The starting parameters are the optimal circuit configuration in the lossless case. The result of the re-optimization is an improvement in the symmetric effective squeezing of the heralded state at the cost of decreasing the success rate, as shown in Fig.\ \ref{fig:re-opt}. The extent of this trade-off can be fine-tuned by including the success probability in the cost function. 
\section{Summary and outlook}
\label{sec:summary}
In this work, we have extended the LCoG simulation framework by merging it with the coherent state decomposition, creating a new connection between the Gaussian formalism and the Fock basis representation. The extended LCoG framework uses the minimal number of resources for the simulation of mixed non-Gaussian states. With this framework, the simulation and optimization of CV circuits that employ a combination of general-dyne measurements as well as PNRDs (including pseudo-PNRDs) are possible without having to change the representation. Our formalism is particularly fast at simulating circuits with a cascade topology; the circuit can be decomposed into sub-circuits with an entanglement operation between two modes, followed by a measurement on one of the modes. After the partial measurement, the number of Gaussians in the single-mode state can be reduced using methods from Sec.\ \ref{sec:state_reduce} which are connected to the stellar rank of the state. This topology is also favorable for simulations in the Fock basis, but Gaussian operations increase the spread over the photon-number basis. In the LCoG framework, Gaussian operations merely change the covariance matrix and the displacement vectors. 

Additionally, we have shown that evaluating the quality of quantum states via the quantum fidelity or characteristic function can be done straightforwardly in the LCoG formalism by computing a sum over exponential functions. It is also possible to obtain the partial derivatives of these functions. We have derived the gradients of the quality with respect to a parameter of a Gaussian gate, which can be obtained via linear algebra starting with the partial derivatives of the covariance matrix and displacement vector in the LCoG framework.

We have showcased the extended LCoG framework by optimizing Gaussian Boson Sampling circuits for the preparation of non-Gaussian states using only a covariance matrix, displacement vectors, and coefficients as well as their partial derivatives. Beyond this application, we expect our methods to be particularly useful for the simulation of non-Gaussian operations on a subset of a larger entangled Gaussian state. For example, our framework can be used to study the spread of non-Gaussianity when part of a cluster state has been measured with a PNRD.

The coherent state decomposition of Fock state superpositions also enables the study of the evolution of non-Gaussian states with finite stellar rank under Gaussian channels, as well as sampling the outcomes of Gaussian measurements using methods from \cite{bourassa_fast_2021}, or simulating their teleportation through a cluster state. For example, the simulation of breeding \cite{vasconcelos_all-optical_2010, weigand_generating_2018, hahn_deterministic_2022,zheng_gaussian_2023} of mixed non-Gaussian states could be performed with our methodology. All of the methods described here are available in an open source library. 

Future areas of research could be to  investigate the viability of tensor network based methods, which are particularly powerful at simulating CV circuits with a shallow entanglement structure. It is not straightforward whether the coherent state decomposition can be connected to matrix product states, since the coherent states do not form an orthonormal basis. While the LCoG framework can include many sources of experimental noise, we have not included phase-noise. It can be modeled by randomly rotating the state according to a Gaussian distribution, which can be simulated shot-by-shot. However, in order to simulate the ensemble, one ends up with many rotations, which is computationally inefficient. There could also be other non-Gaussian operations that can be decomposed into a LCoG, apart from PNRDs. 

In conclusion, we expect the extended LCoG framework to be a useful addition to the growing number of tools for simulating CV quantum optical circuits, with an emphasis on its capability to perform mixed state simulation and interoperate between photon and generaldyne detectors.

\section{Code availability}
The simulation code is written in Python and available at \url{https://github.com/qpit/lcg_plus}. The repository contains several Jupyter Notebooks with tutorials. Scripts for generating the  optimization data and figures can be found at \url{https://github.com/qpit/gbs-optimization}. The optimization were run on the Niflheim super-computing cluster at DTU Physics.

\section{Acknowledgments}
OS thanks Eli Bourassa and Pavithran Iyer for discussions about the linear combination of Gaussians formalism. OS also thanks Ben Baragiola for bringing the coherent state decomposition to our attention. The authors also thank Niklas Budinger for his feedback on the manuscript, which greatly improved its presentation. \\

The authors acknowledge support from the Danish National Research Foundation (bigQ, no.\ DNRF0142), the European Union's Horizon Europe CLUSTEC project (no.\ 101080173), the Innovation Fund Denmark PhotoQ project (no.\ 1063-00046B), the European Research Council ClusterQ project (no.\ 101055224, ERC 2021-ADG), and the Novo Nordisk Foundation Data Science Research Infrastructure 2022 grant (no.\ NNF22OC0078009).\\

The authors would also like to acknowledge the use of several open source libraries \cite{harris_array_2020,virtanen_scipy_2020,hunter_matplotlib_2007,kluyver_jupyter_2016}

\newpage

\bibliography{main} 
\onecolumngrid
\appendix
\section{Homodyne measurements}
\label{app:homodyne}
For multi-mode partial homodyne measurements, that is projections onto the rotated multimode position eigenstate, on a multi-mode Gaussian state, Eq.\ \eqref{eq:sigmaA_gauss} and Eq.\ \eqref{eq:muA_gauss} have the special form \cite{brask_gaussian_2022},
\[
&\vb*{\sigma}_{A}'=\vb*{\sigma}_{A}-\vb*{\sigma}_{AB}(\vb*{\Pi}\vb*{\sigma}_{B}\vb*{\Pi})^{-1}\vb*{\sigma}_{AB}^T, \\
&\vb*{\mu}_{A}'(\vb*{m})=\vb*{\mu}_{A}+\vb*{\sigma}_{AB}(\vb*{\Pi}\vb*{\sigma}_{B}\vb*{\Pi})^{-1}(\vb*{u}-\vb*{\mu}_{B}). \label{eq:homodyne}
\]
For a measurement of the position quadratures, $\vb*{\Pi}=\bigoplus_{i\in B}\begin{pmatrix}
1 & 0  \\
0 & 0
\end{pmatrix}$, $(\vb*{\Pi}\vb*{\sigma}_{B}\vb*{\Pi})^{-1}=\left[\vb*{\sigma}_{B,x}\right]^{-1}\vb*{\Pi}$ where $\vb*{\mu}_{B,x}=\langle \hat{x}_i\rangle$ and $\vb*{\sigma}_{B,x}=\langle \{\hat{x}_{i}-\langle \hat{x}_{i} \rangle  , \hat{x}_{j}-\langle \hat{x}_{j} \rangle \} \rangle$ are the elements belonging to the position operator for $i\in B$, $\vb*{m}$ is a vector over the $x$-measurement outcomes, and $p(\vb*{m})=G_{\vb*{\sigma}_{B,x},\vb*{\mu}_{B,x}}(\vb*{m})$. Similarly, for $p$ measurements, we take the elements of the covariance matrix and displacement vector that belong to the momentum quadrature of the measured subsystem $B$, $\vb*{\Pi}=\bigoplus_{i\in B}\begin{pmatrix}
0 & 0  \\
0 & 1
\end{pmatrix}$, $(\vb*{\Pi}\vb*{\sigma}_{B}\vb*{\Pi})^{-1}=\left[\vb*{\sigma}_{B,p}\right]^{-1}\vb*{\Pi}$ where $\vb*{\mu}_{B,p}=\langle \hat{p}_i\rangle$ and $\vb*{\sigma}_{B,p}=\langle \{\hat{p}_{i}-\langle \hat{p}_{i} \rangle  , \hat{p}_{j}-\langle \hat{p}_{j} \rangle \} \rangle$ for $i\in B$, $\vb*{m}$ is a vector over the $p$-measurement outcomes, and $p(\vb*{m})=G_{\vb*{\sigma}_{B,p},\vb*{\mu}_{B,p}}(\vb*{m})$. For alternating $x/p$ measurements, one can mix and match. \\

For partial homodyne measurements of states in the LCoG formalism in Eq.\ \eqref{eq:Wigner_sumofG}, each covariance matrix in Eq.\ \eqref{eq:sigmaA_sum_of_G}, displacement vector in Eq.\ \eqref{eq:muA_sum_of_G}, and Gaussian overlap factor in Eq.\ \eqref{eq:gamma_sum_of_G} has this special form \cite{brask_gaussian_2022}, where the quadratures can be selected with the $\vb*{\Pi}$ matrix. The number of Gaussian does not increase when performing generaldyne measurements, but one must remember to multiply each Gaussian by the Gaussian weight factor $w_k$.

\section{Writing the pPNRD POVM as a linear combination of thermal states}
\label{app:ppnrd}
We provide a short derivation of Eq.\ \eqref{eq:pPNRD_POVM}. Using that the no click POVM element for a single on/off detector with quantum efficiency $\eta$ has the photon-number representation \cite{ferraro_gaussian_2005},
\[
\hat{\Pi}_{0}(\eta)=\sum_{n=0}^\infty (1-\eta)^n \ketbra{n},
\]
and the definition of a thermal state with occupation number $\Bar{n}$ is
\[
\hat{\rho}_{\text{th}}(\bar{n})=\frac{1}{\bar{n}+1}\sum_{n=0}^\infty \left( \frac{\bar{n}}{\bar{n}+1} \right)^n \ketbra{n}.
\]
By setting $\eta=(\bar{n}+1)^{-1}$, we can relate $\hat{\Pi}_{0}$ to a thermal state via
\[
\hat{\Pi}_{0}(\bar{n})=\sum_{n=0}^\infty\left( \frac{\bar{n}}{\bar{n}+1} \right)^n \ketbra{n} = \underbrace{ (\bar{n}+1) }_{ \eta^{-1} }\hat{\rho}_{\text{th}}(\bar{n}). 
\]
For quantum efficiency $\eta=1$, $\bar{n}=0$, and the thermal state only has support on the vacuum $\ketbra{0}$. For $\eta=0$, $\bar{n}=\infty$ and $\hat{\Pi}_0(\Bar{n})$ becomes the identity $\hat{\mathds{I}}=\sum_{n=0}^\infty \ketbra{n}$ (the zero in the numerator and denominator cancel). \\

The POVM element for a pPNRD consisting of $M$ click detectors is given by \cite{paul_photon_1996, provaznik_benchmarking_2020},
\[
\hat{\Pi}_{k,M} = \sum_{n=0}^\infty p(k|n)\ketbra{n},
\]
where $p(k|n)$ is the probability of getting $k$ clicks given $n$ photons,
\[
p(k|n)=\begin{pmatrix}
M \\
k
\end{pmatrix}
\frac{1}{M^n}\sum_{i=0}^k \begin{pmatrix}
k \\
i
\end{pmatrix}(-1)^i (k-i)^n \quad\text{for}\quad k\leq n\leq M,
\]
and zero otherwise. Using $\eta_{i}=(M-k+i)/M$ so that $1-\eta_{i}=(k-i)/M$, we can rewrite the pPNRD POVM as a linear combination of single-mode off detections with quantum efficiency $\eta_{i}$ (having reordered the sums),
\[
\hat{\Pi}_{k,M}= \begin{pmatrix}
M \\ k
\end{pmatrix}\sum_{i=0}^k\begin{pmatrix} 
k \\ i
\end{pmatrix}(-1)^{i}\sum_{n=0}^\infty(1-\eta_{i})^n\ketbra{n}
\]
From this, we can straightforwardly relate the pPNRD POVM to a linear combination of thermal states,
\[
\hat{\Pi}_{k,M}= \begin{pmatrix}
M \\ k
\end{pmatrix}\sum_{i=0}^k  \begin{pmatrix}
k \\ i
\end{pmatrix}(-1)^i \eta_{i}^{-1}\hat{\rho}_{\text{th}}\left( \bar{n}_{i}=\frac{1-\eta_{i}}{\eta_{i}} \right)
\]
For $(k=M,i=0)$ we recover the $\eta_{i}=0$ limit (identity), while the $(k,i=k)$ results in the $\eta_{i}=1$ limit (vacuum).

\section{Finding the coherent state outer product parameters $\alpha$, $\beta$, $d$ from a complex Gaussian} \label{app:G_to_coherent}
Given a state $\hat{\sigma}$ with Wigner function,
\[
W_{\hat{\sigma}}(\vb*{q})&=\sum_{k\in\mathcal{M}} e^{w_k} G_{\vb*{\mu}_{k}, \frac{\hbar }{2}\mathds{1}_{2}}(\vb*{q}),
\]
the state can be written as a linear combination of coherent state outer products,  \[
\hat{\sigma}&=\sum_{k\in\mathcal{M}}  e^{w_k-d_k}\ketbra{\alpha_k}{\beta_k}
\]
through the following conversion: For $\vb*{\mu}_k=\begin{pmatrix}
\nu_{x}+i\omega_{x} \\
\nu_{p}+i \omega_{p}
\end{pmatrix}$, where $\vb*{\nu}=(\nu_x,\nu_p)^T$ is the real part of the displacement vector, and $\vb*{\omega}=(\omega_x,\omega_p)^T$ is the imaginary part, we can find $\alpha_k$, $\beta_k$, and $d_k$ to be
\[
\alpha_k=\frac{1}{2}\left(\nu_{x}-\omega_{p}+i(\nu_{p}+\omega_{x})\right), \\
\beta_k=\frac{1}{2}\left(\nu_{x}+\omega_{p} +i(\nu_{p}-\omega_{x}) \right), \\
%2\mathrm{Im}(\alpha)=\mu_{p}^{r}+\mu_{x}^{i} \\
%2\mathrm{Im}(\beta)=\mu_{p}^{r}-\mu_{x}^{i} \\
d_k= -\frac{1}{2}\omega_{x}^2 -\frac{1}{2}\omega_{p}^2+\frac{i}{2}(\nu_{p}\omega_{p}+\nu_{x}\omega_{x}).\label{eq:mu_to_alpha_coherent}
\]
For $\vb*{\mu}_k^*=\begin{pmatrix}
\nu_{x}-i\omega_{x} \\
\nu_{p}-i \omega_{p}
\end{pmatrix}$, $\alpha_k$ and $\beta_k$ are swapped, and $d_k\mapsto d_k^*$.

\section{Converting a non-convex sum of coherent state outer products into the Fock basis}
\label{app:coherent_to_fock}
Let $\hat{\sigma}$ be a single mode state in the form of Eq.\ \eqref{eq:dm_sqz_gauss_superpos},
\[
\hat{\sigma}=\hat{U}\sum_{j\in\mathcal{J}} \frac{c_j}{d_j} \ketbra{\alpha_j}{\beta_j}\hat{U}^\dagger
\]
where $\hat{U}$ is a general Gaussian operator and $\mathcal{J}$ is a set of indices. If we sandwich $\hat{\sigma}$ between two identity operators, written in Fock basis, $\hat{\mathds{I}}=\sum_n\ketbra{n}$, we obtain the Fock representation of the core state. 

\[
\hat{\sigma}
% &=\sum_{k} \frac{w_{k}}{d_{j}}\ketbra{\alpha_k}{\beta_k} \\
&=\hat{U}\sum_{mn} \underbrace{\sum_{j} \frac{c_j}{d_{j}}\langle  m|\alpha_{j}\rangle \langle \beta_{j}|n \rangle}_{\rho_{mn}} \ketbra{m}{n} \hat{U}^\dagger
\]
where $\rho_{mn}$ is evaluated to be
\[
\rho_{mn} &= \sum_{j} \frac{c_{j}}{d_{j}} e^{-\frac{1}{2}\lvert \alpha_{j} \rvert ^2-\frac{1}{2}\lvert \beta_j\rvert^2 }\frac{\alpha_{j}^m (\beta_{j}^{*})^n}{\sqrt{ m! n! }} \label{eq:coherent_to_fock_dm}
% &=\frac{1}{\sqrt{ n! m! }}\left[\sum_{j=1}^{k_1} w_j e^{-\lvert \alpha_{j} \rvert ^2}\alpha_{j}^n (\alpha_{j}^{*})^m\right.\\
% &\left.+\sum_{j=k_1+1}^{k_2}e^{-\frac{1}{2}\lvert \alpha_{j} \rvert ^2-\frac{1}{2}\lvert \beta_j \rvert^2 }\left(\frac{w_j}{d_j}\alpha_j^n(\beta_j^*)^m +\frac{w_j^*}{d_j^*}\beta_j^n(\alpha_j^*)^m\right)\right] \label{eq:coherent_to_fock_dm}
\]
If the $mn$ sum is truncated, Eq.\ (36) from \cite{marshall_simulation_2023} can be used to convert back into the coherent state decomposition with potentially fewer Gaussians.

\section{The reduced linear combination of Gaussians (red-LCoG) formalism }\label{app:reduced_formalism}
Let a state $\hat{\rho}$ be expressed in the form of Eq.\ \eqref{eq:wigner_sum_of_G_reduced},
\[
W_{\hat{\rho}}(\vb*{q})=\sum_{k\in\mathcal{M}_R} e^{a_k} G_{\vb*{\mu}_k,\vb*{\sigma}_{k}}(\vb*{q}) + \Re\left[\sum_{l\in\mathcal{M}_I} e^{a_{l}}G_{\vb*{\mu}_{l},\vb*{\sigma}_{l}}(\vb*{q})\right]\label{eq:red_gauss_state},
\]
where the Gaussians are separated into those with $\mathcal{M}_R=\{k|\Im(a_k)=0\} \cap \{k|\Im(\vb*{\mu}_k)=0\} $ real coefficients and means (the covariance matrix is assumed here to be real), and those with $\mathcal{M}_I=\{k\notin \mathcal{M}_R\}$ complex coefficients and means. The normalization constant can be computed by simply taking the real part of the sum over the weights, 
\[
\mathcal{N}=\Re\left[\sum_{k\in\mathcal{M}_R}a^{a_k}+\sum_{l\in\mathcal{M}_I}a^{a_l}\right].
\]
Gaussian operations on this state can be performed as usual on the displacement vectors and covariance matrices. However, for other operations or figures of merit, simply taking the real part of the result will not be correct.\\

Below, we outline how to treat the complex terms when performing measurements, calculating overlaps and evaluating the characteristic function. 
\subsection{Measurements}\label{app:reduced_measurements}
 The state in Eq.\ \eqref{eq:red_gauss_state} is measured with a POVM element $\hat{\Pi}_B$ which is also in the red-LCoG form,  
\[
W_{\hat{\Pi}_B}(\vb*{q}_B)=\sum_{n\in\mathcal{N}_R} e^{b_n} G_{\vb*{\nu}_{n},\vb*{\omega}_{n}}(\vb*{q}_B) +  \Re\left[\sum_{m\in\mathcal{N}_I}e^{b_m}G_{\vb*{\nu}_{m},\vb*{\omega}_{m}}(\vb*{q}_B)\label{eq:red_gauss_povm}\right],
\]
with $\mathcal{N}_R=\{n|\Im(b_n)=0\} \cap \{n|\Im(\vb*{\nu}_n)=0\}$ and $\mathcal{N}_I=\{n\notin \mathcal{N}_R\}$. The post-selected state consists of a number of partial phase-space integrals, 
\[
W_{\hat{\rho}_{A}'}(\vb*{q}_A)&\propto \sum_{k\in\mathcal{M}_R}\sum_{n\in\mathcal{N}_R}e^{a_k+b_n+\gamma_{kn}}G_{\vb*{\mu}_{kn,A}'\vb*{\sigma}_{kn,A}'}(\vb*{q}_{A})\nonumber\\
&+\underbrace{ \int \dd{\vb*{q}_{B}}\sum_{k\in\mathcal{M}_R}e^{a_k}G_{\vb*{\mu}_{k},\vb*{\sigma}_{k}}(\vb*{q})\mathrm{Re}\left[ \sum_{m\in\mathcal{N}_I}e^{b_m}G_{\vb*{\nu}_{m},\vb*{\omega}_{m}}(\vb*{q}_{B}) \right] }_{ T_{1} }\nonumber \\
&+\underbrace{ \int \dd{\vb*{q}_{B}}\mathrm{Re}\left[\sum_{l\in\mathcal{M}_I}e^{a_l}G_{\vb*{\mu}_{l},\vb*{\sigma}_{l}}(\vb*{q}) \right] \sum_{n\in\mathcal{N}_R}e^{b_n}G_{\vb*{\nu}_{n},\vb*{\omega}_{n}}(\vb*{q}_{B}) }_{ T_{2} }\nonumber \\
&+\underbrace{ \int \dd{\vb*{q}_{B}}\mathrm{Re}\left[\sum_{l\in\mathcal{M}_I}e^{a_l}G_{\vb*{\mu}_{l},\vb*{\sigma}_{l}}(\vb*{q}) \right]\mathrm{Re}\left[\sum_{m\in\mathcal{N}_I}e^{b_m}G_{\vb*{\nu}_{m},\vb*{\omega}_{m}}(\vb*{q}_{B}) \right] }_{ T_{3} }.
\]
The first integral is already computed using the transformation rules from Sec \ref{sec:sumofG}. The rest of the integrals can be computed by 
expanding the real elements in terms of their complex conjugates,
\[
\Re\left[\sum_{k\in \mathcal{M}}e^{a_{k}}G_{\vb*{\mu}_{k},\vb*{\sigma}_{k}}(\vb*{q})\right] = \frac{1}{2}\sum_{k\in \mathcal{M}}\left(e^{a_{k}}G_{\vb*{\mu}_{k},\vb*{\sigma}_{k}}(\vb*{q}) + e^{a_{k}^*}G_{\vb*{\mu}_{k}^*,\vb*{\sigma}_{k}^*}(\vb*{q})\right).
\]
This allows us to use the transformation rules of partial Gaussian measurements to calculate the transformed Gaussian,
\[
T_{1}=\mathrm{Re}\left[ \sum_{k\in\mathcal{M}_R}\sum_{m\in\mathcal{N}_I} e^{a_k+b_m+\gamma_{km}}G_{\vb*{\mu}_{km,A}',\vb*{\sigma}_{km,A}'}(\vb*{q}_{A})\right],
\]
\[
T_{2}=\mathrm{Re}\left[\sum_{l\in\mathcal{M}_I}\sum_{n\in\mathcal{N}_R} e^{a_l +b_n+\gamma_{ln}}G_{\vb*{\mu}_{ln,A}',\vb*{\sigma}_{ln,A}'}(\vb*{q}_{A})\right],
\]
\[
T_{3}&=\mathrm{Re}\left[\sum_{l\in\mathcal{M}_I}\sum_{m\in\mathcal{N}_I}\left(e^{a_l+b_m+\gamma_{lm}}G_{\vb*{\mu}_{lm,A}',\vb*{\sigma}_{lm,A}'}(\vb*{q}_{A}) +e^{a_l+b_{\tilde{m}} +\gamma_{l\tilde{m}}} G_{\vb*{\mu}_{l\tilde{m}, A}',\vb*{\sigma}_{l\tilde{m}, A}'}(\vb*{q}_{A})\right)\right].
\]
where the tilde accent, $\tilde{m}$ on the second term means to use the complex conjugate of the $m$ terms, i.e.
\[
\gamma_{l \tilde{m}}=-\frac{1}{2}(\vb*{\mu}_{l,B}-\vb*{\nu}_{m}^*)^T(\vb*{\sigma}_{l,B}+\vb*{\omega}_{m}^*)^{-1}(\vb*{\mu}_{l,B}-\vb*{\nu}_{m}^*)-\frac{1}{2}\ln(2\pi \det(\vb*{\sigma}_{l,B}+\vb*{\omega}_{m}^*)),
\]
and similarly for $\vb*{\mu}_{l\tilde{m},A}'$ and $\vb*{\sigma}_{l\tilde{m},A}'$. 

If the initial state had $k_1$ real Gaussians and $k_2$ complex Gaussians, and the POVM element had $n_1$ real Gaussians and $n_2$ complex Gaussians, then the projective measurement would result a state with $k_1n_1$ real Gaussians and $k_1n_2+n_1k_2+2k_2n_2$ complex Gaussians. Giving a total of $\frac{1}{2}k_1n_1(k_1n_1+1)$ Gaussians. 

\subsection{Overlap}\label{app:reduced_overlap}
Similar to the partial measurements, in order to calculate the fidelity or overlap between two states in the red-LGoG form, we must take special care of the complex Gaussian terms. Let the Wigner function of the two states $\hat{\rho}_1$ and $\hat{\rho}_2$ be
\[
W_{\hat{\rho}_1}(\vb*{q})=\sum_{k\in\mathcal{M}_R} e^{a_k} G_{\vb*{\mu}_k,\vb*{\sigma}_{k}}(\vb*{q}) + \Re\left[\sum_{l\in\mathcal{M}_I} e^{a_{l}}G_{\vb*{\mu}_{l},\vb*{\sigma}_{l}}(\vb*{q})\right]\label{eq:red_gauss_state1},\\
W_{\hat{\rho}_2}(\vb*{q})=\sum_{n\in\mathcal{N}_R} e^{b_n} G_{\vb*{\nu}_n,\vb*{\omega}_{n}}(\vb*{q}) + \Re\left[\sum_{m\in\mathcal{N}_I} e^{b_m}G_{\vb*{\nu}_{m},\vb*{\omega}_{m}}(\vb*{q})\right]\label{eq:red_gauss_state2}.
\]
The overlap between $\hat{\rho}_1$ and $\hat{\rho}_2$ requires an additional sum over the overlaps between the complex Gaussians,
\[
\Tr[\hat{\rho}_1\hat{\rho}_2]&=\sum_{k\in\mathcal{M}_R}\sum_{n\in\mathcal{N}_R}e^{a_k+b_n}G_{\vb*{\mu}_{k},\vb*{\sigma}_{k}+\vb*{\omega}_{n}}(\vb*{\nu}_{n})\nonumber\\
&+\mathrm{Re}\left[ \sum_{k\in\mathcal{M}_R}\sum_{m\in\mathcal{N}_I}e^{a_k+b_m}G_{\vb*{\mu}_{k},\vb*{\sigma}_{k}+\vb*{\omega}_{m}}(\vb*{\nu}_{m}) \right]\nonumber\\
&+\mathrm{Re}\left[\sum_{l\in\mathcal{M}_I}\sum_{n\in\mathcal{N}_R}e^{a_l + b_n}G_{\vb*{\mu}_{l},\vb*{\sigma}_{k}+\vb*{\omega}_{n}}(\vb*{\nu}_{n}) \right]\nonumber\\
&+\mathrm{Re}\left[] \sum_{l\in\mathcal{M}_I}\sum_{m\in\mathcal{N}_I}e^{a_l+b_m}G_{\vb*{\mu}_{l},\vb*{\sigma}_{l}+\vb*{\omega}_{m}}(\vb*{\nu}_{m}) \right]\nonumber\\
&+\mathrm{Re}\left[\sum_{l\in\mathcal{M}_I}\sum_{m\in\mathcal{N}_I}e^{a_l+b_m^*}G_{\vb*{\mu}_{l},\vb*{\sigma}_{l}+\vb*{\omega}_{m}^*}(\vb*{\nu}_{m}^*) \right].
\]
\subsection{Characteristic function}
The characteristic function must be computed via the full representation, i.e. including the complex conjugates of the means and displacement vectors,
\[
\chi_{\hat{\rho}}(\vb*{\alpha})&=\sum_{k\in\mathcal{M}_{R}}e^{w_{k}+i\vb*{\mu}_{k}^T \vb*{\Omega}\vb*{\alpha}-\frac{1}{2}\vb*{\alpha}^T\vb*{\Omega}\vb*{\sigma}_{k}\vb*{\Omega}^T\vb*{\alpha}}\nonumber\\
&+\sum_{l\in\mathcal{M}_{I}}e^{-\frac{1}{2}\vb*{\alpha}^T\vb*{\Omega}\vb*{\sigma}_{l}\vb*{\Omega}^T\vb*{\alpha}}\left(e^{w_{l}+i\vb*{\mu}_{l}^T \vb*{\Omega}\vb*{\alpha}}+e^{w_{l}^*+i\vb*{\mu}_{l}^{*T} \vb*{\Omega}\vb*{\alpha}}\right)
\]

\section{Photon-number moments of LCoGs}
\label{photon_number_moments}
The first and second order photon-number moment of a single-mode Gaussian state with displacement vector $\vb*{\mu}$ and covariance matrix $\vb*{\sigma}$ are given by \cite{dodonov_multidimensional_1994},
\[
&\mu_n=\langle \hat{n}\rangle = \frac{1}{2\hbar}\left(\Tr[\vb*{\sigma}]+\vb*{\mu}^T\vb*{\mu}\right)-\frac{1}{2}, \label{eq:photon_mean_gauss} \\ 
&\sigma_n=\langle \hat{n}^2 \rangle - \langle \hat{n} \rangle^2 = \frac{1}{2\hbar^2}\left(\Tr[\vb*{\sigma}]^2-2\det[\vb*{\sigma}]+2\vb*{\mu}^T\vb*{\sigma}\vb*{\mu}\right)-\frac{1}{4}.\label{eq:var_photon_gauss}
\]
This can be generalized to states that are LCoGs by weighing the photon number moments of each Gaussian.
\[
\mu_n &= \sum_{k\in\mathcal{K}} e^{w_{k}} \mu_n^{[k]}=\sum_{k\in\mathcal{K}} e^{w_{k}} \frac{1}{2\hbar}\left(\mathrm{Tr}[\vb*{\sigma}_{k}]+\vb*{\mu}_{k}^T \vb*{\mu}_{k}-\hbar\right),\label{eq:photon-number_mean_sumG}\\ 
\sigma_n &= \sum_{k\in\mathcal{K}} e^{w_{k}}\sigma_n^{[k]}=\sum_{k\in\mathcal{K}}e^{w_{k}} \frac{1}{2\hbar^2}\left( \mathrm{Tr}[\vb*{\sigma}_{k}]^2-2\det[\vb*{\sigma}_{k}]+2\vb*{\mu}_{k}^T\vb*{\sigma}_{k}\vb*{\mu}_{k} - \frac{\hbar^2}{2} \right). \label{eq:photon-number_variance_sumG}
\]
When calculating photon-number moments of approximate Fock states in the coherent state decomposition in Eq.\ \eqref{eq:wigner_fock_approx}, we obtain a peculiar result; the photon-number variance of this state is equal to the mean photon-number, i.e. $\mu_n=n$ and $\sigma_n=n$, which is a characteristic of coherent states. The photon-number variance of a Fock state should of course be zero. This deviation in the expected behavior can be explained by the nature of the coherent state decomposition. The approximate Fock state has non-zero amplitude on $\ket{n}$, $\ket{2n+1}$, $\ket{3n+2}$ and so on \cite{marshall_simulation_2023}. The small radius $\epsilon$ of the coherent state ring reduces the amplitude of the higher order Fock contributions, making it a good approximation of $\ket{n}$. However, the infinite Fock basis expansion for finite $\epsilon$ (and in principle infinite stellar rank, similar to cat states) is still captured by the photon-number variance. 

Consequently, the photon-number variance calculated via Eq.\ \eqref{eq:photon-number_variance_sumG} for states in the coherent state decomposition should be interpreted as an estimate. When the photon-number variance is used to find the minimum number of coherent states needed to express a mixed state in the coherent decomposition, we are no longer certain of the error we are making. However, in practice by taking a large number of standard deviations of the photon-number in Chebyshev's inequality, we can make sure that the error is small. 

\section{Updating the gradients of the log-weights, means and covariance matrix after a measurement}\label{app:partial}

We begin with the partial derivative of the covariance matrix $\frac{\partial \vb*{\sigma}}{\partial \phi}$ and displacement vector $\frac{\partial\vb*{\mu}}{\partial \phi}$ of a multi-mode Gaussian state $W_{\hat{\rho}}(\vb*{q})=G_{\vb*{\mu},\vb*{\sigma}}(\vb*{q})$ with respect to a parameter $\phi$ of a Gaussian unitary. Let's furthermore partition the system into subsystem $B$ (containing the first mode) and subsystem $A$ (the rest of the modes), 
\[
\vb*{\sigma}=\begin{pmatrix}
\vb*{\sigma}_{A} & \vb*{\sigma}_{BA}  \\
\vb*{\sigma}_{AB}^T & \vb*{\sigma}_{B}
\end{pmatrix}, \quad
\frac{\partial \vb*{\sigma}}{\partial \phi} = \begin{pmatrix}
\frac{\partial\vb*{\sigma}_{A}}{\partial \phi} & \frac{\partial\vb*{\sigma}_{AB}}{\partial \phi} \\
\frac{\partial\vb*{\sigma}_{BA}}{\partial\phi} & \frac{\partial\vb*{\sigma}_{B}}{\partial\phi}
\end{pmatrix},
\quad \vb*{\mu}=\begin{pmatrix}
\vb*{\mu}_{A} \\
\vb*{\mu}_{B}
\end{pmatrix}, \quad \frac{\partial\vb*{\mu}}{\partial \phi}=\begin{pmatrix}
\frac{\partial\vb*{\mu}_{A}}{\partial \phi}  \\
\frac{\partial\vb*{\mu}_{B}}{\partial \phi}
\end{pmatrix}.
\]
When subsystem $B$ is measured with a POVM element in the form of Eq.\ \eqref{eq:wigner_fock_approx}, 
\[
W_{\hat{\Pi}_B}(\vb*{q}_B)=\sum_{j\in\mathcal{J}_1}e^{w_j}G_{\vb*{\nu}_j,\frac{\hbar}{2}\mathds{1}}(\vb*{q}_B).
\]
The rest of the system is transformed according to Eq.\ \eqref{eq:post_select_fock_Wigner},
\[
W_{\hat{\rho}_A'}(\vb*{q}_A)=\sum_{j\in\mathcal{J}_1}e^{w_j+\gamma_{j}}
G_{\vb*{\tilde{\mu}}_{j},\vb*{\tilde{\sigma}}}(\vb*{q}_A).
\]

The partial derivative of the Gaussian overlap factor from Eq.\ \eqref{eq:gamma_GBS} is

\[
\frac{\partial \gamma_{j}}{\partial \phi} &= \frac{1}{2}\frac{\partial \vb*{\mu}_{B}^T}{\partial \phi}(\vb*{\sigma}_{B}+\mathds{1})^{-1}(\vb*{\nu}_{j}-\vb*{\mu}_{B})\nonumber\\
&+\frac{1}{2}(\vb*{\nu}_j-\vb*{\mu}_{B})^T(\vb*{\sigma}_{B}+\mathds{1})^{-1}\frac{\partial \vb*{\mu}_{B}}{\partial \phi}\nonumber\\
&-\frac{1}{2}(\vb*{\nu}_{j}-\vb*{\mu}_{B})^T \frac{\partial (\vb*{\sigma}_{B}+\mathds{1})^{-1}}{\partial \phi}(\vb*{\nu}_{j}-\vb*{\mu}_{B})\nonumber\\
&-\frac{1}{2}\Tr[(\vb*{\sigma}_B+\mathds{1})^{-1}\frac{\partial\vb*{\sigma}_B}{\partial\phi}].\label{eq:partial_weight}
\]
The last term is the partial derivative of $\delta=-\frac{1}{2}\ln(2\pi\det(\vb*{\sigma}_B+\mathds{1}))$, found using Jacobi's formula for derivatives of determinants. The partial derivative of the transformed displacement vector in Eq.\ \eqref{eq:mu_GBS}, is 
\[
\frac{\partial \vb*{\tilde{\mu}}_{j}}{\partial \phi}&=\frac{\partial \vb*{\mu}_A}{\partial \phi}+\frac{\partial \vb*{\sigma}_{AB}}{\partial \phi}(\vb*{\sigma}_{B}+\mathds{1})^{-1}(\vb*{\nu}_{j}-\vb*{\mu}_{B}) \nonumber\\
&+ \vb*{\sigma}_{AB} \frac{\partial (\vb*{\sigma}_{B}+\mathds{1})^{-1}}{\partial \phi} (\vb*{\nu}_{j}-\vb*{\mu}_{B}) -\vb*{\sigma}_{AB}(\vb*{\sigma}_{B}+\mathds{1})^{-1} \frac{\partial \vb*{\mu}_{B}}{\partial \phi}.\label{eq:partial_mean}
\]
The partial derivative of the transformed covariance matrix in Eq.\ \eqref{eq:sigma_GBS} is
\[
\frac{\partial \vb*{\tilde{\sigma}}}{\partial \phi}&= \frac{\partial \vb*{\sigma}_{A}}{\partial \phi}-\frac{\partial \vb*{\sigma}_{AB}}{\partial \phi}(\vb*{\sigma}_{B}+\mathds{1})^{-1} \vb*{\sigma}_{BA}\nonumber\\
&-\vb*{\sigma}_{AB} \frac{\partial(\vb*{\sigma}_{B}+\mathds{1})^{-1}}{\partial \phi} \vb*{\sigma}_{BA}-\vb*{\sigma}_{AB}(\vb*{\sigma}_{B}+\mathds{1})^{-1} \frac{\partial \vb*{\sigma}_{BA}}{\partial \phi}.\label{eq:partial_cov}
\]
The derivative of the matrix inverse is 
\[
\frac{\partial(\vb*{\sigma}_{B}+\mathds{1})^{-1}}{\partial \phi}=-(\vb*{\sigma}_{B}+\mathds{1})^{-1} \frac{\partial \vb*{\sigma}_{B}}{\partial \phi}(\vb*{\sigma}_{B}+\mathds{1})^{-1}.
\]
For the subsequent measurement of the second mode with a POVM element of the same form $W_{\hat{\Pi}_{D}}(\vb*{q}_{D})=\sum_{k\in\mathcal{J}_2}e^{w_k}G_{\vb*{\nu}_k,\frac{\hbar}{2}\mathds{1}}(\vb*{q}_{D})$ the partial derivatives $\frac{\partial \vb*{\tilde{\sigma}}}{\partial \phi}$ and $\frac{\partial \vb*{\tilde{\mu}}_{j}}{\partial \phi}$ can also be decomposed into subsystem $D$ (the second mode) and subsystem $C$ (the remaining modes),
\[
\tilde{\vb*{\sigma}}=\begin{pmatrix}
\tilde{\vb*{\sigma}}_{C} & \tilde{\vb*{\sigma}}_{CD} \\
\tilde{\vb*{\sigma}}_{CD'}^T & \tilde{\vb*{\sigma}}_{D}
\end{pmatrix},\quad 
\frac{\partial \tilde{\vb*{\sigma}}}{\partial \phi} = \begin{pmatrix}
\frac{\partial\tilde{\vb*{\sigma}}_{C}}{\partial \phi} & \frac{\partial\tilde{\vb*{\sigma}}_{CD}}{\partial \phi} \\
\frac{\partial\tilde{\vb*{\sigma}}_{CD}^T}{\partial\phi} & \frac{\partial\tilde{\vb*{\sigma}}_{D}}{\partial\phi},
\end{pmatrix},\quad 
\tilde{\vb*{\mu}}_{j}=\begin{pmatrix}
\tilde{\vb*{\mu}}_{j,C} \\
\tilde{\vb*{\mu}}_{j,D}
\end{pmatrix}, \quad \frac{\partial\tilde{\vb*{\mu}}_j}{\partial \phi}=\begin{pmatrix}
\frac{\partial\tilde{\vb*{\mu}}_{j,C}}{\partial \phi}  \\
\frac{\partial\tilde{\vb*{\mu}}_{j,D}}{\partial \phi}
\end{pmatrix}\label{eq:partial_CD_partition}.
\]
The post-selected state becomes
\[
W_{\hat{\rho}_C'}(\vb*{q}_C)=\sum_{j\in\mathcal{J}_1}\sum_{k\in\mathcal{J}_2}e^{w_j+\gamma_{j}+w_k+\gamma_k}
G_{\vb*{\tilde{\tilde{\mu}}}_{j,k},\vb*{\tilde{\tilde{\sigma}}}}(\vb*{q}_C).
\]
The new partial derivatives $\frac{\partial \gamma_{k}}{\partial \phi}$,  $\frac{\partial \vb*{\tilde{\tilde{\mu}}}_{j,k}}{\partial \phi}$, and $\frac{\partial \vb*{\tilde{\tilde{\sigma}}}}{\partial \phi}$ can be computed from Eq.\ \eqref{eq:partial_weight}, \eqref{eq:partial_mean}, \eqref{eq:partial_cov} from the matrix elements of Eq.\ \eqref{eq:partial_CD_partition}. With this prescription, the partial derivatives can be computed mode-by-mode (measurement-by-measurement) until there is only one mode left. 

\subsection{Partial derivative of the overlap}

\[
\frac{\partial\gamma_{kl}}{\partial \phi} = &-\frac{1}{2}\frac{\partial\vb*{\mu}_{k}}{\partial \phi}(\vb*{\sigma}_{k}+\vb*{\omega}_{l})^{-1}(\vb*{\mu}_{k}-\vb*{\nu}_{l})\nonumber\\ 
&-\frac{1}{2}(\vb*{\mu}_{k}-\vb*{\nu}_{l})^T  \frac{\partial (\vb*{\sigma}_{k}+\vb*{\omega}_{l})^{-1}}{\partial \phi}(\vb*{\mu}_{k}-\vb*{\nu}_{l}) \nonumber\\
&-\frac{1}{2} (\vb*{\mu}_{k}-\vb*{\nu}_{l})^T(\vb*{\sigma}_{k}+\vb*{\omega}_{l})^{-1} \frac{\partial \vb*{\mu}_{k}}{\partial \phi} \label{eq:partial_overlap_gamma}.
\]
 The derivative of the matrix inverse is 
\[
\frac{\partial(\vb*{\sigma}_k+\vb*{\omega}_l)^{-1}}{\partial \phi}=-(\vb*{\sigma}_k+\vb*{\omega}_l)^{-1} \frac{\partial \vb*{\sigma}_k}{\partial \phi}(\vb*{\sigma}_k+\vb*{\omega}_l)^{-1}
\]
\[
\frac{\partial\delta_{kl}}{\partial \phi} = -\frac{1}{2}\mathrm{Tr}(\vb*{\sigma}_{k}+\vb*{\omega}_{l}) \frac{\partial \vb*{\sigma}_{k}}{\partial \phi} \label{eq:partial_overlap_delta}.
\]

\section{Partial derivatives of Gaussian operations}
Unless specified, all operations have null displacement, $\vb*{d}=\vb*{0}$.
\subsection{Single-mode squeezing}
The squeezing operation, $\hat{S}(\zeta)=\exp\left[ \frac{1}{2} (\zeta^* \hat{a}^2-\zeta \hat{a}^{2\dagger})\right]$ where $\zeta=r e^{i\phi}$, has symplectic matrix
\[
\vb*{S}(r,\phi)=\cosh r \mathds{1}_{2} - \sinh r \vb*{S}_{\phi},
\]
where
\[
\vb*{S}_{\phi}=\begin{pmatrix}
\cos \phi & \sin \phi \\ \sin \phi & -\cos \phi
\end{pmatrix}\label{eq:symplectic_Sphi}.
\]
The partial derivatives with respect to the squeezing magnitude and angle are,
\[
 \frac{\partial\vb*{S}}{\partial r}=\sinh r\mathds{1}_{2} - \cosh r\vb*{S}_{\phi},
\]
\[
\frac{\partial\vb*{S}}{\partial \phi}=-\sinh r \frac{\partial\vb*{S}_{\phi}}{\partial \phi}, \quad\text{where}\quad\frac{\partial\vb*{S}_{\phi}}{\partial \phi}=\begin{pmatrix}
-\sin \phi & \cos \phi \\
\cos \phi & \sin \phi
\end{pmatrix}.
\]
\subsection{Two-mode squeezing}
The two-mode squeezing operator,
$\hat{S}(\zeta)=\exp[\zeta^* \hat{a}_{1}\hat{a}_{2}-\zeta \hat{a}_{1}^{\dagger} \hat{a}_{2}^{\dagger}]$ where $\zeta = re^{i\phi}$ has symplectic matrix
\[
\vb*{S}(r,\phi)=\begin{pmatrix}
\cosh r \mathds{1}_{2} & -\sinh r\vb*{S}_{\phi} \\
-\sinh r\vb*{S}_{\phi} & \cosh r \mathds{1}_{2}
\end{pmatrix},
\]
where $\vb*{S}_\phi$ is given in Eq.\ \eqref{eq:symplectic_Sphi}. The partial derivatives are
\[
\frac{\partial\vb*{S}}{\partial r} = \begin{pmatrix}
\sinh r \mathds{1}_{2} & -\cosh r \vb*{S}_{\phi} \\
-\cosh r \vb*{S}_{\phi} & \sinh r \mathds{1}_{2}
\end{pmatrix},
\]
\[
\frac{\partial\vb*{S}}{\partial \phi} = \begin{pmatrix}
0_{2,2} & -\sinh r \frac{\partial\vb*{S}_{\phi}}{\partial \phi}  \\
-\sinh r \frac{\partial\vb*{S}_{\phi}}{\partial \phi} & 0_{2,2}
\end{pmatrix},
\]
where $0_{2,2}$ is the $2\times2$ null matrix.
\subsection{Beamsplitter}
Following the strawberryfields convention,
the beamsplitter is defined as $\hat{B}(\theta ,\phi) =\exp[\theta(e^{i\phi} \hat{a}_{1}\hat{a}_{2}^{\dagger}-e^{-i\phi}\hat{a}^{\dagger}_{1}\hat{a}_{2})]
$, where $t=\cos\theta$ is the transmission amplitude and $r=e^{i\phi}\sin\theta$ is the reflection amplitude. The quadrature operators transform under the following symplectic matrix,
\[
\vb*{B}(\theta,\phi)=\begin{pmatrix}
\cos \theta\mathds{1}_{2} & -\sin \theta \vb*{S}_{\phi}^T \\
\sin \theta \vb*{S}_{\phi} & \cos \theta \mathds{1}_{2}
\end{pmatrix},
\]
where $\vb*{S}_\phi$ is the symplectic matrix of the anticlockwise rotation operation $\hat{R}(\phi)=\exp[i\phi\hat{a}^\dagger \hat{a}]$,
\[
\vb*{S}_{\phi}=\begin{pmatrix}
\cos \phi & -\sin \phi \\
\sin \phi & \cos \phi
\end{pmatrix} \quad\text{and}\quad \frac{\partial\vb*{S}_{\phi}}{\partial \phi} = \begin{pmatrix}
-\sin \phi & -\cos \phi \\
\cos \phi &- \sin \phi
\end{pmatrix},
\]
\[
\frac{\partial\vb*{B}(\theta,\phi)}{\partial \theta} = \begin{pmatrix}
-\sin \theta\mathds{1}_{2} & -\cos \theta \vb*{S}_{\phi}^T \\
\cos \theta \vb*{S}_{\phi} &-\sin \theta \mathds{1}_{2}
\end{pmatrix}
\]
\[
\frac{\partial\vb*{B}(\theta,\phi)}{\partial \phi}=\begin{pmatrix}
0_{2,2} & -\sin \theta\frac{\partial\vb*{S}_{\phi}^T}{\partial \phi} \\
\sin \theta\frac{ \partial\vb*{S}_{\phi}}{\partial \phi} & 0_{2,2}
\end{pmatrix}.
\]
\subsection{Displacement}
$\hat{D}(\alpha)=\exp[\alpha \hat{a}^\dagger-\alpha^* \hat{a}]$ where $\alpha = re^{i\phi}$. transforms with $\vb*{S}=\mathds{1}_2$, so $\frac{\partial\vb*{S}}{\partial r}=\frac{\partial\vb*{S}}{\partial \phi}=0_{2,2}$. The partial derivative of the displacement vector is 
\[
\vb*{d}=\sqrt{2\hbar} \begin{pmatrix}
    \Re\alpha \\ \Im\alpha
\end{pmatrix}=\sqrt{2\hbar}\begin{pmatrix}
    r\cos\phi \\ r\sin\phi
\end{pmatrix},\quad
\frac{\partial\vb*{d}}{\partial r}=\sqrt{ 2\hbar }\begin{pmatrix}
\cos \phi \\
\sin \phi
\end{pmatrix},\quad
\frac{\partial\vb*{d}}{\partial \phi}=\sqrt{ 2\hbar }\begin{pmatrix}
-r\sin \phi \\
r\cos \phi
\end{pmatrix}.
\]

\end{document}